\documentclass[fontsize=10pt, paper=a4]{scrartcl}
\usepackage[T1]{fontenc}
\usepackage{lmodern}
\usepackage[a4paper, left=2.5cm, right=2.5cm, top=2.5cm, bottom=2.5cm]{geometry}
\usepackage[svgnames]{xcolor}
\usepackage[utf8]{inputenc}
\usepackage{appendix}
\usepackage[colorlinks=true,allcolors=blue]{hyperref}
\usepackage{booktabs}
\usepackage{graphicx}
\usepackage{float}
\graphicspath{ {./2-Figures/} }
\usepackage{amsmath}

\DeclareSectionCommand[%
  style=section,
  level=4,
  toclevel=4,
  indent=0pt,
  beforeskip=-3.25ex plus -1ex  minus -.2ex,
  afterskip=1.5ex plus .2ex,
  counterwithin=subsubsection,
]{subsubsubsection}
\RedeclareSectionCommand[level=5, toclevel=5, counterwithin=subsubsubsection]{paragraph}
\RedeclareSectionCommand[level=6, toclevel=6]{subparagraph}
\addtocounter{tocdepth}{1}

\usepackage{soul}
\newcommand{\highlight}[1]{{\sffamily\bfseries #1}}

\newcommand{\email}[1]{\href{mailto:#1}{\texttt{#1}}}

\setkomafont{title}{\vspace{20mm}\sffamily\LARGE\bfseries}
\setkomafont{subtitle}{\vspace{2mm}\normalfont\sffamily\Large\bfseries}
\setkomafont{author}{\vspace{10mm}\normalfont\sffamily\large}

\usepackage{biblatex}
\title{Bridging Systems}
\subtitle{Open Problems for Countering Destructive Divisiveness across\\ Ranking, Recommenders, and Governance}

\author{
    \textbf{Aviv Ovadya}\footnote{Aviv Ovadya is an affiliate at the Berkman Klein Center for Internet \& Society at Harvard University, the Centre for the Governance of AI, and the Thoughtful Technology Project. This work began while he was a Technology and Public Purpose Fellow at the Harvard Kennedy School's Belfer Center (2021-2022).} \\ 
    Harvard University \\ 
    \email{aviv@aviv.me} \and 
    \textbf{Luke Thorburn}\footnote{Luke Thorburn is a researcher in the UKRI Centre for Doctoral Training in Safe and Trusted AI at King's College London.} \\
    King's College London \\
    \email{mail@lukethorburn.com}
}
\date{}

\newcommand{\faClipboard}{\raisebox{-2pt}{\includegraphics[width=9pt]{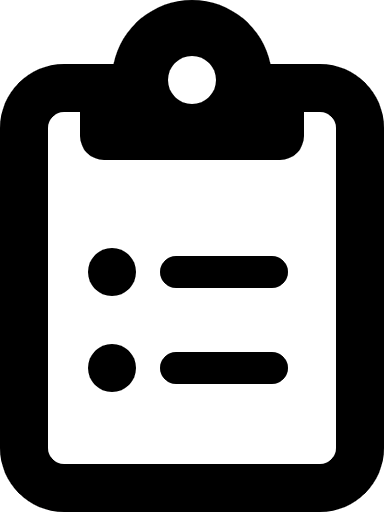}}\hspace{4pt}}

\newcommand{\faSort}{\raisebox{-1.5pt}{\includegraphics[width=11pt]{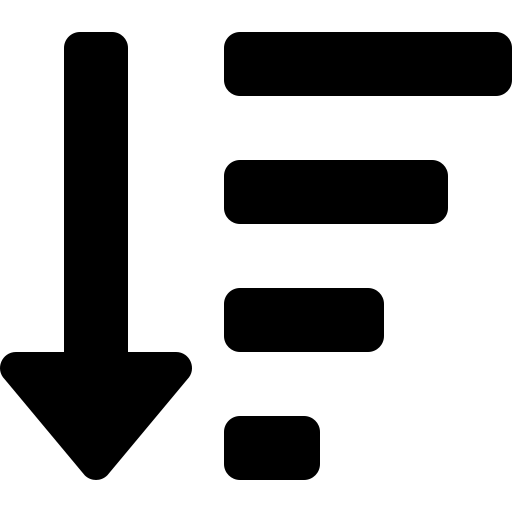}}\hspace{3pt}}
\newcommand{\faComments}{\raisebox{-1.5pt}{\includegraphics[width=11pt]{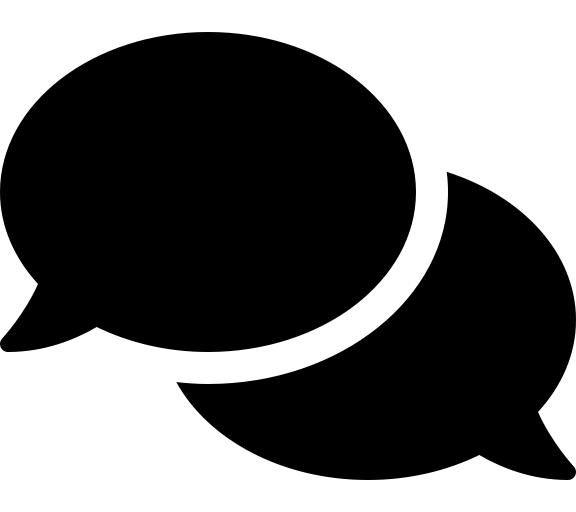}}\hspace{3pt}}
\newcommand{\faWarning}{\raisebox{-1.5pt}{\includegraphics[width=11pt]{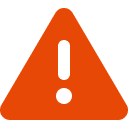}}}

\definecolor{question}{HTML}{E0F2F1}
\newcounter{question}
\newenvironment{questions}[1][]{
        \refstepcounter{question}%
        \vspace{2mm}\begin{tcolorbox}[
            detach title,
            before upper={\tcbtitle\par\medskip},
            title={ \textsf{\textbf{Open Questions}}: #1 \hfill \textsf{ \emph{Set~\thequestion}}},
            coltitle=black,
            colback=question,
            colframe=white,
            left=1mm,
            right=1mm,
            top=1mm,
            bottom=1mm,
            boxsep=2mm,
            boxrule=0mm,
            arc=0mm
        ]
        
        \setdefaultleftmargin{21pt}{16pt}{}{}{}{}
        \vspace{-3mm}
        \begin{enumerate}
    }{
        \end{enumerate}
        \end{tcolorbox}\vspace{2mm}
    }
    
\definecolor{example}{HTML}{F1F8E9}
\newcounter{example}

\newenvironment{exampleResponse}[1][]{
        
        \refstepcounter{example}%
        \vspace{2mm}\begin{tcolorbox}[
            detach title,
            before upper={\tcbtitle\par\medskip},
            title={ \faClipboard \ \textsf{\textbf{Collective Response System}: #1} \hfill \textsf{ \emph{Example~\theexample}}},
            coltitle=black,
            colback=example,
            colframe=white,
            left=1mm,
            right=1mm,
            top=1mm,
            bottom=1mm,
            boxsep=2mm,
            boxrule=0mm,
            arc=0mm
        ]
    }{
        \end{tcolorbox}\vspace{2mm}
    }
    
\newenvironment{exampleRecommender}[1][]{
        \refstepcounter{example}%
        \vspace{2mm}\begin{tcolorbox}[
            detach title,
            before upper={\tcbtitle\par\medskip},
            title={ \faSort \ \textsf{\textbf{Recommender System}: #1} \hfill \textsf{\emph{Example~\theexample}} },
            coltitle=black,
            colback=example,
            colframe=white,
            left=1mm,
            right=1mm,
            top=1mm,
            bottom=1mm,
            boxsep=2mm,
            boxrule=0mm,
            arc=0mm
        ]
    }{
        \end{tcolorbox}\vspace{2mm}
    }
    
\newenvironment{exampleFacilitation}[1][]{
        \refstepcounter{example}%
        \vspace{2mm}\begin{tcolorbox}[
            detach title,
            before upper={\tcbtitle\par\medskip},
            title = { \faComments \ \textsf{\textbf{Human Facilitation}: #1} \hfill \textsf{\emph{Example~\theexample}} },
            coltitle=black,
            colback=example,
            colframe=white,
            left=1mm,
            right=1mm,
            top=1mm,
            bottom=1mm,
            boxsep=2mm,
            boxrule=0mm,
            arc=0mm
        ]
    }{
        \end{tcolorbox}\vspace{2mm}
    }

\definecolor{glossary}{HTML}{AAAAAA}

\usepackage{footnote}
\usepackage{etoolbox}
\BeforeBeginEnvironment{exampleResponse}{\savenotes}
\BeforeBeginEnvironment{exampleRecommender}{\savenotes}
\BeforeBeginEnvironment{exampleFacilitation}{\savenotes}
\AfterEndEnvironment{exampleResponse}{\spewnotes}
\AfterEndEnvironment{exampleRecommender}{\spewnotes}
\AfterEndEnvironment{exampleFacilitation}{\spewnotes}
\usepackage{amssymb} %
\usepackage{marginnote} %

\newcommand{\note}[1]{ 
}

\usepackage{tcolorbox}
\tcbuselibrary{many}
\usepackage[hybrid]{markdown}

\usepackage[colorinlistoftodos]{todonotes}

\newcommand{\todoNow}[1]{ 
}
\newcommand{\todoPhase}[1]{ 
}
\newcommand{\feedback}[1]{ 
}

\newcommand{\todoCite}[1]{ %
}
\newcommand{\todoLater}[1]{ 
}

\newcommand{\newterm}[1]{\emph{#1}}
\newcommand{\term}[1]{\texttt{#1}}
\newcommand{\variable}[1]{\texttt{#1}}

\usepackage{paralist}
\usepackage{enumitem}

\usepackage{fancyhdr}
\definecolor{darkred}{HTML}{D50000}
\begin{document}

\maketitle

\begin{abstract}

    \noindent
    Divisiveness appears to be increasing in much of the world, leading to concern about political violence and a decreasing capacity to collaboratively address large-scale societal challenges. In this paper we aim to articulate an interdisciplinary research and practice area focused on what we call \textit{bridging systems}: systems that increase mutual understanding and trust across divides, creating space for productive conflict, deliberation, or cooperation. We give examples of bridging systems across three domains: recommender systems on social media, collective response systems, and human-facilitated group deliberation. We argue that these examples can be more meaningfully understood as processes for  \textit{attention allocation} (as opposed to ``content distribution'' or ``amplification'') and develop a corresponding framework to explore similarities---and opportunities for bridging---across these seemingly disparate domains. We focus particularly on the potential of \textit{bridging-based ranking} to bring the benefits of offline bridging into spaces that are already governed by algorithms. Throughout, we suggest research directions that could improve our capacity to incorporate bridging into a world increasingly mediated by algorithms and artificial intelligence.

\end{abstract}

\begin{center}
    \vspace{14mm}
    \sffamily\large
    This document is a draft. A revised version will be published with the\\ Knight First Amendment Institute at Columbia University.
\end{center}

\clearpage

\tableofcontents
\clearpage

\section{Introduction}

    \begin{quotation}
        Imagine a platform that gave people status not for clever takedowns of political opponents but for producing content with bipartisan appeal. ... Instead of boosting content that is controversial or divisive, such a platform could improve the rank of messages that resonate with different audiences simultaneously.\\ 
        \vspace{-6mm}
        \begin{flushright}
            \highlight{Chris Bail}, \emph{Breaking the Social Media Prism} \cite{bail2021}
        \end{flushright}
    \end{quotation}

\paragraph{Division impacts cooperation and conflict.}
    
    We face compounding global challenges including climate change, pandemics, and transformative artificial intelligence, all of which are likely to require significant cooperation to navigate. %
    At the same time, there is significant public concern around increasing societal division \cite{svolik2019,mccoy2019} and the resulting increase in \newterm{destructive conflict} \cite{deutsch1969ConflictsProductiveDestructive}---which can increase the likelihood of large-scale political violence \cite{walter2022HowCivilWars}. 
    Destructive conflict and the violence that can result from it not only harms innumerable lives directly---it may also make addressing those global challenges exceedingly difficult \cite{ovadya2021HoldingPlatformsAccountable}. 

\paragraph{The systems that allocate our attention can impact division.}
    
    Increases in societal division may be related to the incentives of the systems which guide people's attention---what we call \newterm{attention allocators}. We are all attention allocators in that we have some agency over how we allocate our own limited attention. But before we can even choose among what to attend to, \newterm{upstream} attention allocators such as the recommender systems on social media platforms, search engines, news media, and even human facilitation have already done much of that allocation for us. From a firehose of potential information, they choose a much smaller set of items for us to attend to in our limited time.
    
    In this paper we focus on these upstream attention allocators, as they shape the incentives of our attention economy by directing attention to some kinds of behavior over others \cite{williams2018}. Because attention can translate to money and power, such systems help determine what kinds of behaviors are rewarded in many spheres of life. %

\paragraph{Systems that directly reward attention may have ``bias toward division''.}

    Many attention allocators reward behavior that seeks to maximize attention toward themselves. Recommender systems, for example, largely seek to maximize measures of attention (what is commonly referred to as \emph{engagement-based ranking}), and most news media entities need to attract attention for their work out of financial necessity. Similarly, politicians often aim to attract attention in order to win elections, and anecdotally those who craft online messages for politicians report needing to use inflammatory language in order to be competitive \cite{stray2023}. Incentives such as these reward \newterm{engagement-bait}---content and ideas intended to generate engagement, which are often misleadingly sensational and hyperbolic. In practice, engagement-bait can crowd out good faith efforts to communicate across divides, decrease understanding and trust among people of diverse viewpoints, and thus increase division \cite{lorenzspreen2022,brady2021}.

    Two common ways of defining division are \newterm{ideological polarization} (differences in policy positions) and \newterm{affective polarization} (emotional dislike of those from the other party). Measurements of affective polarization, for example, show increases in many parts of the world \cite{boxell2022}.\footnote{Recent work has also moved towards using multiple measures of division, including ``resistance to cross-partisan collaboration'', ``resistance to interpersonal contact with outpartisans'', and ``willingness to use violent tactics against outpartisans'' \cite{voelkel2022}.} This increase appears to be correlated with the increasing adoption of ubiquitous digital communication, but the extent to which this relationship is causal remains contested \cite{lorenzspreen2022,haidt2021}.

\paragraph{This paper explores how to incentivize \emph{bridging}.} 
    
    The goal of bridging is to increase mutual understanding and trust across divides, creating space for productive conflict, deliberation, or cooperation.
    Every system that involves human attention---from social media recommendations, to search engine ranking, to governance processes---will, to some extent, reward or punish bridging. We explore two core questions: \emph{What do systems that reward bridging look like? How might they be designed?}

    Our goal is to clearly articulate an interdisciplinary research area that can inform the development and accelerate the adoption of bridging systems across domains. In other words, just as there are research areas devoted to other biases we aim to articulate a research agenda focused on overcoming the ``bias toward division.''

\paragraph{Bridging is \emph{not} about eliminating conflict or creating homogeneity.}

\begin{figure}[h]
        \includegraphics[width=\textwidth]{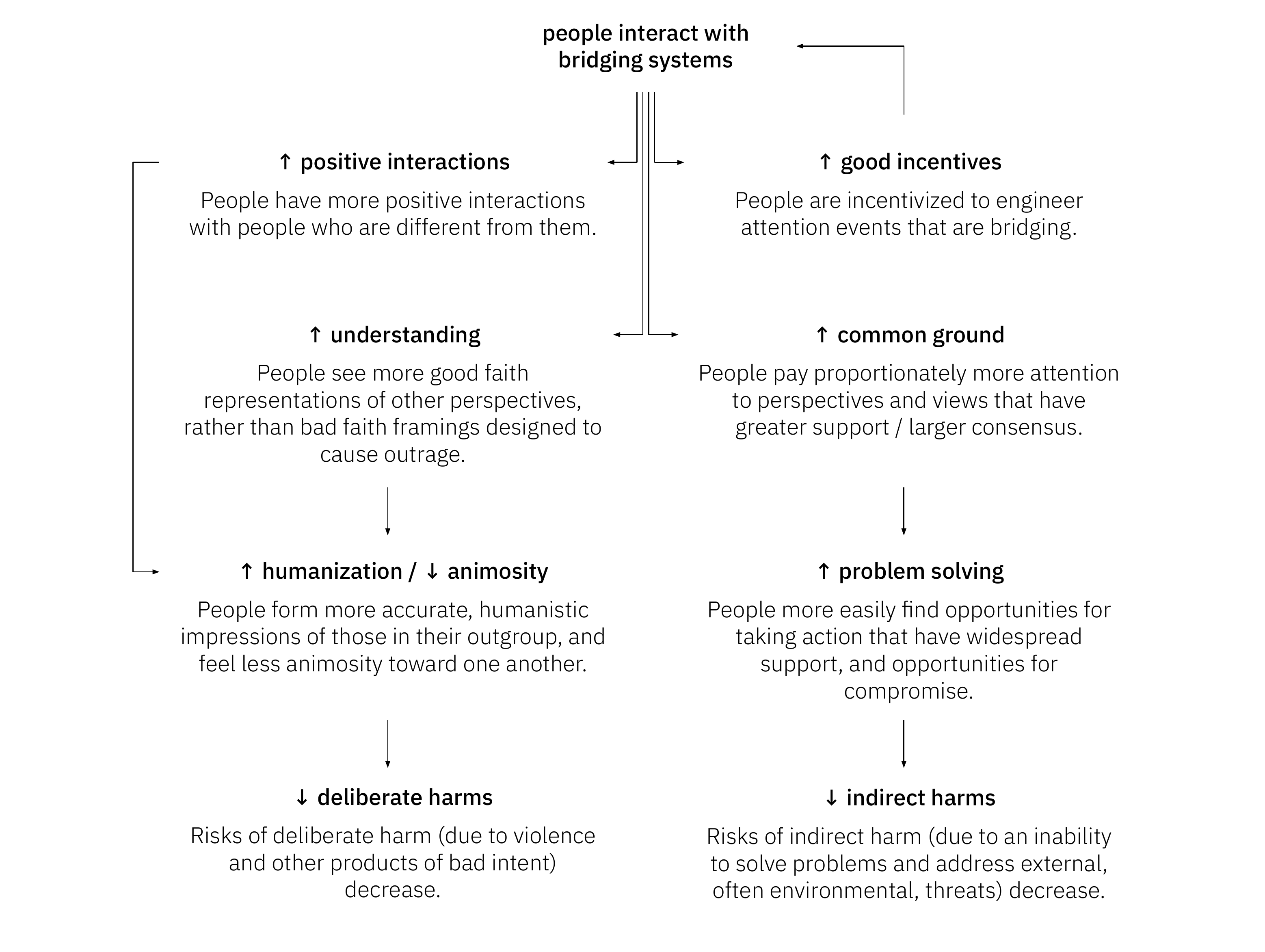}
        \vspace{-8mm}
        \caption{\highlight{A causal loop diagram illustrating how bridging systems might impact society.} The goal of this diagram is not to make strong, precise claims about causality, but simply to provide intuition on how a proliferation of bridging systems could have important and beneficial societal consequences and reduce both deliberate and indirect harms. Significant work is required to determine which causal relations hold (including those not drawn on this diagram), and under what conditions.}
        \label{fig:theory-of-change}
    \end{figure}

    The goal is to check the default tendencies of many environments (including much of social media), which can potentially push us toward extremes when combined with our psychological predilections.\footnote{An economics framing of the ultimate goal might describe it as countering the ``externality of incentives for divisive behavior'' by ``intentionally subsidizing bridging''.} 
    As articulated by \textcite{stray2021b}, the intent is ``conflict transformation'' \cite{lederach2015,botes2003}: not to remove divisions or interfere with the substance of civic debates, but to ``[make] conflict better in some way''. In other words, to mitigate the risks of ``high conflict'' \cite{ripley2021} whilst supporting ``healthy'' or ``constructive'' conflict. When we use the terms bridging, or reducing division, we are thus \emph{not} referring to ``making everyone believe the same things''. Those terms are instead shorthand for ``enabling mutual understanding and respect across divides''---in other words, supporting pluralism \cite{weyl2022}.\footnote{
    For example, a system might bridge divides by facilitating understanding that one's existing values and beliefs are much more similar than expected to those of other people, closing what has been called the ``perception gap'' \cite{yudkin2019}. In this way, beliefs about \emph{other people's beliefs} are changing, while personal beliefs are staying fixed.} Figure~\ref{fig:theory-of-change} provides a speculative causal loop diagram \cite{coleman2019} illustrating the potential ideal impacts of bridging systems.

\paragraph{Moving from engagement-based ranking to \emph{bridging-based ranking}.} 

\begin{figure}[ht]
        \includegraphics[width=\textwidth]{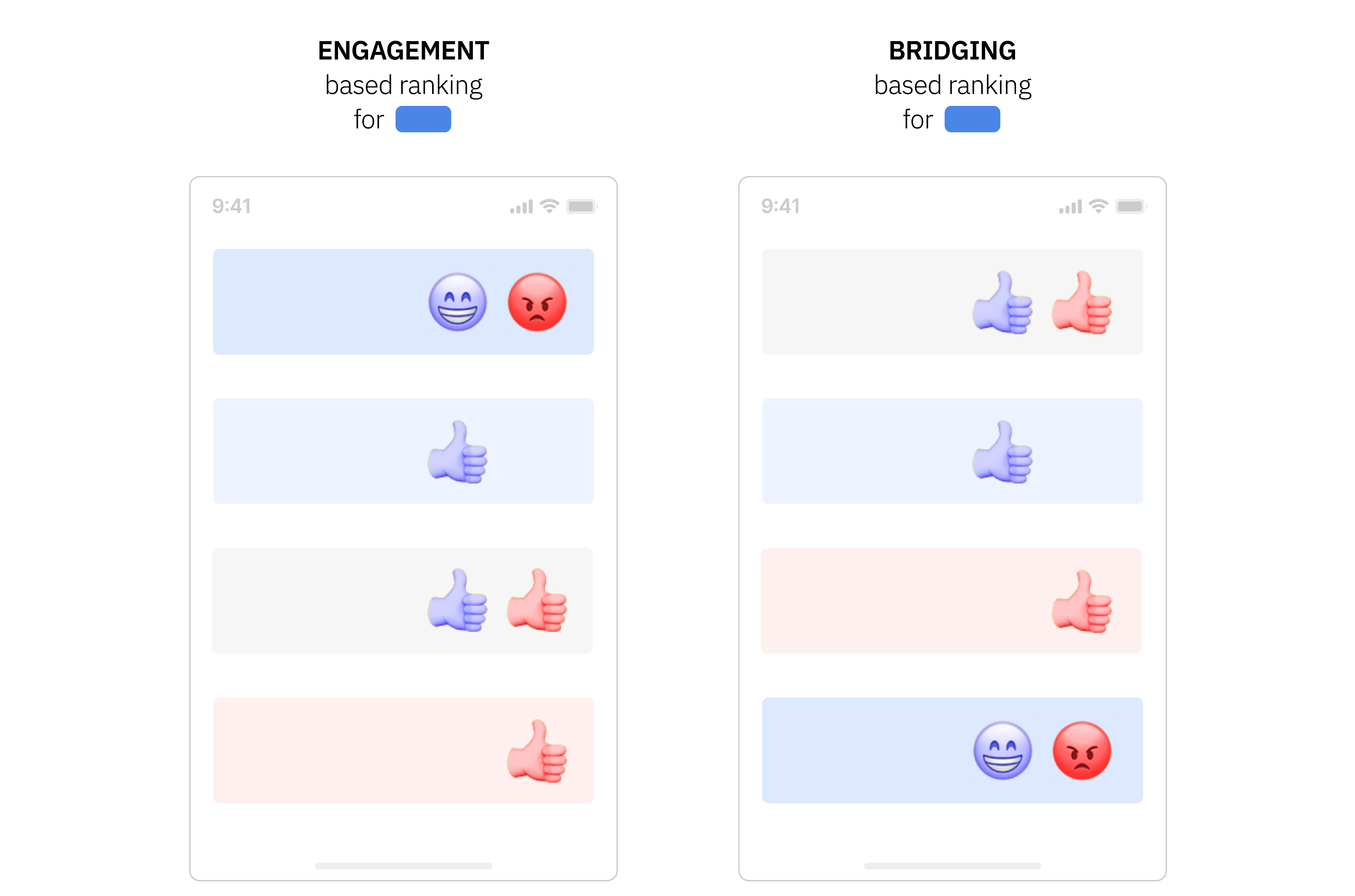}
        \caption{\highlight{A simple example of bridging-based ranking.} Under engagement-based ranking, posts are ranked highly for a user if they are liked by similar users, regardless of the stance of dissimilar users. In contrast, under bridging-based ranking---formalized here using a diverse approval motif (Section \ref{sec:signal-examples})---the post ranked highest is liked by both parties, and the lowest-ranked post is the most divisive. This illustration conveys important intuition, but note that it oversimplifies both the sophistication of status quo engagement-based ranking and the potential sophistication of bridging-based ranking.}
        \label{fig:ranking}
    \end{figure}

    By definition, optimizing more for bridging means optimizing less for engagement, but the extent to which these two goals are in tension is an open question. It may be possible to include bridging impacts within the objective function of a recommender system without undermining financial sustainability. Figure~\ref{fig:ranking} gives a simple example of what such bridging-based ranking \cite{ovadya2022} might look like in the context of a recommender system on a social media platform.

\subsection{Background}

    Throughout this paper we will use examples from three domains to illustrate the concept of a bridging system: (1) recommender systems on social media platforms, (2) software for supporting large-scale civic discussion, and (3) facilitated in-person deliberation and mediation. Each of these domains is briefly introduced below.
    
    \begin{exampleRecommender}[Overview]
        A recommender system is an algorithm that selects which items of content, from a large pool of available items, should be shown to a user \cite{aggarwal2016}. Recommender systems are commonly used on social media platforms---prominent examples include the algorithms behind the Twitter timeline, the Facebook feed, and the YouTube homepage.
        
        \quad Their implementations vary, but most recommender systems on social media operate using the same basic logic: they ``optimize for engagement'', selecting the items of content that are most likely to elicit clicks, likes, reactions, comments, reshares, and other behaviors that the platform can measure~\cite{thorburn2022}. Such behaviors are often an effective proxy by which to select what people want \cite{thorburn2022a}, but can also incentivize the creation of content that is misleading, sensational, outrageous, or addictive \cite{bengani2022}.
    \end{exampleRecommender} \vspace{-5mm}

    \begin{exampleResponse}[Overview]
        Governments, civil society organizations, and others that act on behalf of large groups of people often need to elicit the views of their constituents and stakeholders. A variety of tools for coordinating such large-scale dialogues have been created, including some that explicitly seek to elevate common ground.
        In this document, we call these \newterm{collective response systems} \cite{ovadya2023} and focus on two illustrative examples.
        
        \quad The first example, \href{https://web.archive.org/web/20140125190913/https://yourview.org.au/forums/1-Fairfax-YourView-Election-Forum/}{YourView} \cite{yourview2014}, was a not-for-profit collective response system that operated in Australia during the lead-up to the 2013 federal election \cite{short2012,vangelder2012,vangelder2012a}. YourView provided concise explanations of policy proposals, under which participants could contribute comments for or against the proposal, as well as vote on the comments of other participants, and on the issue overall. From this voting data, YourView derived two measures of public support for the proposal: the raw percentage of participants in favor, and a ``Public Wisdom'' percentage in which the participants with the highest ``credibility score'' were given the most weight.
        
        \quad The second example, \href{https://pol.is/}{Polis} \cite{polis2022}, is an open-source collective response system that has been used to conduct public consultations informing digital policy in Taiwan, among other applications \cite{hsiao2018,small2021}. Polis instances are context-specific forums where participants can contribute comments and vote on the comments of others. From this voting data, Polis clusters participants according to their voting patterns and generates visualizations of the support for each comment among the participants in each cluster.
    \end{exampleResponse} \vspace{-5mm}
    
    \begin{exampleFacilitation}[Overview]
        
        While there is a very broad range of potential examples of human facilitation, here we focus specifically on mini-publics \cite{escobar2017}. They involve the convening of a diverse group of people to deliberate over a particular policy issue, guided by impartial facilitators. The group is selected through sortition, providing a representative random sample (roughly analogous to the selection of jury candidates in the criminal justice system, though more similar in process to representative polling). Increasingly, such mini-publics are being used to make progress on contested political issues, including abortion laws in Ireland \cite{courant2021} and climate change policy in France \cite{giraudet2022}. There is an emerging profession of facilitators who are skilled at convening and coordinating such groups \cite{dryzek2019,carson2020}. Mini-public facilitation is of particular interest because, unlike many multi-stakeholder forums, participants are not just elites and represent a wide range of personal experiences and viewpoints in a single facilitation environment.
    \end{exampleFacilitation}

    The term ``bridging'' has been used by others. For example, literature on social capital often draws distinctions between three kinds of social ties: bonding, bridging, and linking \cite{szreter2004}. Of these, both bridging and linking are used to refer to connections between individuals who differ in some way (``horizontally'' and ``vertically'', respectively). We use bridging more broadly and generically to describe any process that increases mutual understanding and trust across divides (see Section \ref{sec:bridging-property}), though we agree with the claim that bridging ties are a core part of a connected society \cite{allen2016}. Bridging has also been used by others to describe strategies for connecting across differences during offline, face-to-face interactions \cite{shigeoka2020,powell2022}, the effects of which have been extensively studied in the research area of ``intergroup contact theory'' and found to be quite robustly positive \cite{pettigrew2011}. Many of these strategies may form the basis for formal, quantitative signals used to identify bridging algorithmically (Section \ref{sec:signals}).

    There are a number of existing research efforts that seek to use technology in ways that satisfy our definition of ``bridging'', and on which our work builds. These include work on the use of recommender systems for depolarization \cite{keulenaar2018,stray2021b,badami2017,badami2018,badami2021,zhu2021,celis2019,rastegarpanah2019}, work that seeks to build insights from conflict mediation and peacebuilding communities into algorithmic settings \cite{keulenaar2018,schirch2023,schirch2023a,council2023,stray2023,burgess2022}, and work that seeks to quantify the deliberative quality of an argument or discussion \cite{fournier-tombs2021,vecchi2021}.
    More broadly, the design of bridging systems draws on work from many disciplines that inform our understanding of sociotechnical systems.\footnote{There are also many associated non-academic bodies of knowledge around attention allocators generally and bridging specifically which can be learned from across industry, government, and civil society.} This includes work on:

    \begin{itemize}
        
        \item \emph{articulating what ``good'' public discourse looks like} for attention allocators generally, from domains including political theory, philosophy (e.g., epistemology), economics (e.g., social choice theory), communication, anthropology, sociology, rhetoric, and science \& technology studies.%
        
        \item \emph{the collection of accurate social data}, from domains across the behavioral and social sciences including, quantitative history, measurement theory, and survey design;
        
        \item \emph{modeling social phenomena}, from domains including computational social science, opinion dynamics,
        game theory, and political science;
        
        \item \emph{the design of interfaces and environments}, from domains including human-computer interaction (e.g., data visualization), economics (e.g., mechanism design, choice architecture), science \& technology studies, organizational design, facilitation design, urban planning, and architecture. 

    \end{itemize}
    
    Even this broad set of disciplines is far from comprehensive, and we do not attempt a thorough review of all relevant work here.
    However, in many cases, there are direct precursors to particular bridging system components. We will cite many of these in the examples discussed throughout the paper.

\subsection{Contribution}

    Our goal in this paper is to articulate a research and practice direction around bridging systems. %
    Section \ref{sec:attention allocation-systems} provides a framework for thinking about attention allocators, illustrated using the examples of recommender systems, collective dialogue systems, and human-facilitated deliberations, and defines bridging as a property of such systems (\ref{sec:bridging-property}). Section \ref{sec:data-and-modeling} describes how relationships in a population can be formally modelled, and Sections \ref{sec:signals} and \ref{sec:metrics} give concrete examples of signals and metrics that might be used to instantiate the bridging goal. Considerations for evaluating bridging systems are described in Section \ref{sec:evaluation}, and Section \ref{sec:discussion} discusses the challenges, limitations and risks of bridging. Throughout, we propose intellectually compelling and societally impactful open problems, to help support both cross-domain collaboration and rapid beneficial experimentation and deployment. 

    Depending on their own goals, readers may wish to allocate their attention to particular sections.
    \begin{itemize}
        
        \item \textbf{\sffamily Academics} interested in helping progress our understanding of bridging might focus on the technical discussion in Sections \ref{sec:attention allocation-systems}-\ref{sec:metrics}, and on the blue boxes throughout which list open questions.

        \item \textbf{\sffamily Platform employees} seeking to experiment with bridging should in the first instance focus on the discussion of signals and metrics in Sections \ref{sec:signals} and \ref{sec:metrics}, and the examples given throughout in green boxes.

        \item \textbf{\sffamily Regulators} and \textbf{\sffamily civil society actors} might benefit most from the attention allocation framework in Section \ref{sec:attention allocation-systems}, and the discussion of risks and limitations in Section \ref{sec:discussion}.
    
    \end{itemize}  
    We also provide an accompanying glossary at \href{https://bridging.systems/definitions/}{\url{https://bridging.systems/definitions/}} \cite{definitions2023} containing the concepts we introduce and use throughout the paper.

\clearpage

\section{Attention Allocation Systems}
\label{sec:attention allocation-systems}

    \begin{quotation}
        \noindent If a tree falls in a forest and no one is around to hear it, does it make a sound? \\
        If content is distributed and no attention is paid to it, does it even matter?
    \end{quotation}

    \noindent
    We all have control over how we allocate our personal attention, but we do not do this alone. Out of necessity, much of this allocation work is first partially delegated to upstream social and technological systems. These include our three recurring examples---recommender systems, collective response systems, and human-facilitated mini-publics---but also books, media organizations, search engines, group chats, large language models, and many other systems.

    In this section, we formally define allocation and attention in order to provide a general model of attention allocation systems (or attention allocators). Our goal is to facilitate the translation of insights across domains (for example, from offline human facilitation into algorithmic recommender systems), and to help identify points of intervention, in order to improve attention allocators. Finally, we frame bridging as a property of attention allocation systems.

\paragraph{Why allocation?}

    While it is common to speak of systems for ``content distribution'' or (algorithmic) ``amplification'', we believe that ``attention allocation'' is often a more useful frame and term because it acknowledges that attention (as we will define it) is scarce. In particular, human attention is finite and any given individual will only be able to attend to a small fraction of the content to which they have access. In many situations, economic allocation of attention is quite literally what occurs \cite{hwang2020}, and speaking in terms of allocation avoids the ambiguities of the term ``amplification'' \cite{thorburn2022c}.

\paragraph{What is allocation?}

    Allocation is the filling of \newterm{slots} with \newterm{objects}. A slot is a finite resource, and an object is anything that can consume that resource. In this way, the two terms (\textit{slot} and \textit{object}) are defined circularly, but it should be clear from context which is which. For example, an object might be an item of content on a social media platform, and the corresponding slots might be discrete positions within a recommender feed. Alternately, slots might be continuous intervals of first-person experience, in which case the corresponding objects would be anything that can be attended to.
    
    The simplest form of allocation occurs when one slot is filled with one object. We describe these simple allocations as \newterm{atomic}. Atomic allocations are represented as
    \begin{equation*}
        (\variable{slot},\ \variable{object},\ \variable{properties}).
    \end{equation*}
    The third element, \variable{properties}, is a catch-all for data that describe the nature, qualities, and context of the allocation, and which are formalized differently in different contexts. For example, if the allocation represents a post appearing in a ranked recommender feed, \variable{properties} might contain the context in which the feed is viewed (time of day, type of device), engagement that resulted, and so on. More complex allocations can be described as sets of atomic allocations. For example, an allocation of attention where a thousand people attend to the same object would be represented as the set of atomic allocations describing each individual attending to that object.

    \begin{exampleRecommender}[Allocation]
        Recommender systems (and more broadly, the user interfaces of social media platforms) create atomic allocations each time a particular item of content (an \variable{object}) is used to fill a particular position within a recommender feed (a \variable{slot}). The \variable{properties} include engagement data such as how long the user subsequently paused on the item (dwell time), but also the social context such as reactions and comment counts which were shown alongside the content.
    \end{exampleRecommender} \vspace{-5mm}
    
    \begin{exampleResponse}[Allocation]
        Collective response systems often have two kinds of allocations, both involving the allocation of attention. The first occurs when users are shown items for evaluation (e.g., dialogue, voting, etc.), and the second occurs when the most widely supported items (e.g., the results of the votes) are presented to everyone.
    \end{exampleResponse} \vspace{-5mm}

    \begin{exampleFacilitation}[Allocation]
        In facilitated mini-publics, allocations of attention occur fluidly as people engage in dialogue. Here, the \variable{object} might be an idea being expressed by another person, and the additional \variable{properties} of the allocation include the tone, facial expressions, and body language with which it is communicated.
    \end{exampleFacilitation}

    A \newterm{potential} allocation is one that has not yet happened (and may or may not happen), and a \newterm{realized} allocation is one that has already happened.

\subsection{Allocation}

    An \newterm{allocation system} (or simply \newterm{allocator}) is a process that determines which of many potential allocations will actually occur, taking as input a set of potential allocations and outputting a set of realized allocations.
    
    Allocators consist of an \newterm{allocation process} and, optionally\footnote{The learning process is optional because some systems (e.g., chronological recommender systems) do not require learning.}, a \newterm{learning process}.

\subsubsection{Allocation Process}
\label{sec:allocation-process}

    The allocation process is the core of an allocation system. Its purpose is to determine which of a set of potential allocations to realize---that is, which objects to use to fill a finite set of available slots. The process takes as inputs a set of potential allocations and predicts how each allocation would, if realized, change the state of the world along several dimensions. These predicted impacts are aggregated into a measure of the ``worthiness'' or ``utility'' of each allocation using a normative \newterm{value model}.\footnote{The value model formally defines what it means for an allocation to be ``worthy''. For example, a recommender system might define value to be a weighted sum of different measures of engagement, while a human facilitator will have a qualitative value model that balances the needs of group members to be heard with the need for the group to deliver on its remit.} Because global predictions may not be possible or accurate, local heuristics or other signals are also often used for impact prediction. Finally, the most valuable allocations (according to the value model) are selected and realized.
    
    \begin{figure}[th]
        \centering
        \includegraphics[width=0.9\textwidth]{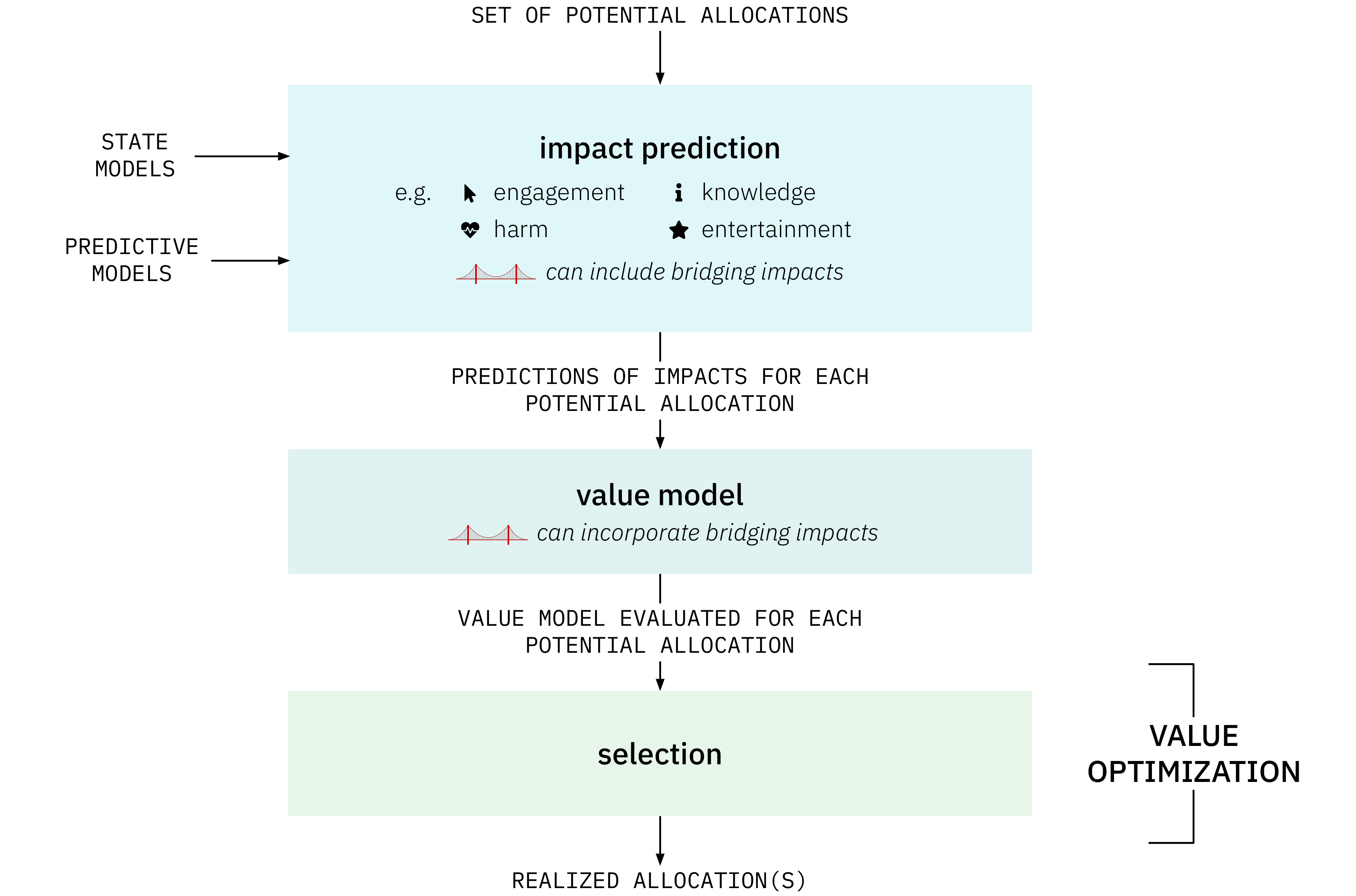}
        \vspace{2mm}
        \caption{\highlight{The allocation process in an allocation system.} A bridge icon indicates where bridging can be incorporated. Not shown are the ways in which the process itself is optimized.}
        \label{fig:allocation-process}
    \end{figure}

\subsubsection{Learning Process}
\label{sec:learning-process}

    The purpose of the learning process is to improve the predictions generated by the allocation process. When a trigger indicates that models need updating, relevant data is retrieved from storage, collected, or elicited from the people interacting with the system\footnote{Going forward, we will simply use ``data collection'' as a catch-all term.}. This data is used to update (or fine-tune) both \newterm{state models}---static descriptions of the current state of the world---and \newterm{predictive models}---models that predict the impacts of allocations, optionally conditioned on the current state. These updated models are then substituted into the allocation process.

    \note{Collection: from external sources / world; Retrieval: from internal sources / database; Elicitation: directly from agents.}

    \begin{figure}[h]
        \centering
        \includegraphics[width=0.9\textwidth]{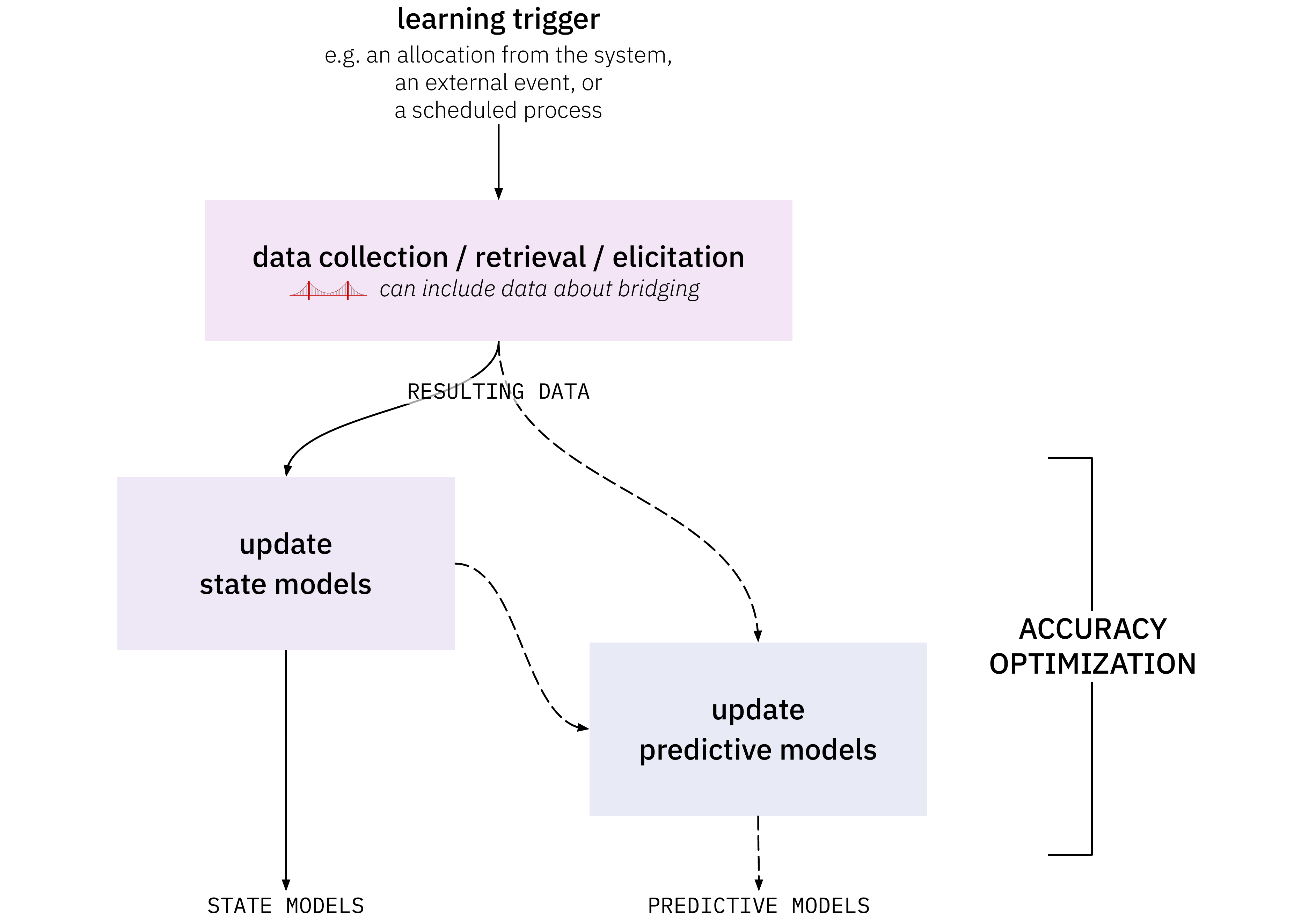}
        \caption{\highlight{The learning process in an allocation system.} A bridge icon indicates where bridging can be incorporated. The dashed lines indicate that many technical systems will update state models after each attention allocation, but update predictive models less frequently. Human facilitators update their implicit state models as they facilitate (noticing how different events impact people), and also update their predictive models slightly as they work, but might do a much larger ``update'' of their predictive models (improving their facilitation skills) during a post-facilitation retrospective. Not shown are implicit inputs such as the previous models, or how the act of data elicitation can itself change the state of the world.}
        \label{fig:learning-process}
    \end{figure}
    
\subsection{Attention Allocation}

    We define \newterm{attention} as the selective processing of information by a system, such that the state of that system might materially change.\footnote{This is an intentionally broad definition, which may be more expansive than much of the current usage across machine learning and psychology.} We use the terms \newterm{attend to} or \newterm{pay attention to} to describe the enacting of that ``potential material change'' to the system---systems that can only create outputs (without any internal changes) are not \newterm{attending}. As an example, when a person (a kind of system) attends to a news article (a kind of information), their beliefs and attitudes may change as a result of them attending to that article. Because people (and other systems) face limitations in their access to information and have only a finite amount of processing capacity, they can benefit from external systems that help allocate that limited attention.
 
    We can now define an \newterm{attention allocation system} (or \newterm{attention allocator}) as an allocation system that is involved in the allocation of attention. Crucially, an attention allocator does not itself need to attend (and many do not attend under our definition, for example a recommender system that has been trained once on existing data and is not updated).

    Our three examples can be viewed as attention allocators, to which people (partially) delegate the allocation of their attention.

    \begin{exampleRecommender}[Attention allocator]
        For a given social media account, every item of content corresponds to a potential atomic allocation or, more accurately, to multiple potential atomic allocations, one for each context and position in which the content could appear in the user interface\footnote{In practice, most items will be removed from consideration during a \newterm{candidate generation} phase, to reduce computational cost \cite{thorburn2022}.}. A set of impacts of each of these potential allocations is predicted. Commonly considered impacts include engagement behaviors (e.g., clicks, comments, shares), whether the user will be entertained (e.g., what they would rate the content), or whether they are likely to be harmed (e.g., whether the content is a financial scam). The predictions are then aggregated using a formal value model \cite{thorburn2022}, and the resulting scores are the primary factor that determines how items are ranked within the recommender feed, influencing which allocations take place. For the most part, recommenders currently direct attention to optimize engagement \cite{thorburn2022}, but could target other goals \cite{stray2021b}.
    \end{exampleRecommender} \vspace{-5mm}

    \begin{exampleResponse}[Attention allocator]
        As mentioned, collective response systems often have two kinds of attention allocation. For example, Polis only presents one item at a time for voting, and it tries to choose which item to show in order to learn as much as possible about the overall structure of the views and perspectives present in a population. Thus the main ``predicted impact'' in the allocation process is the information about participants' views provided by the allocation.
        
        \quad Polis then uses that information to create a non-personalized ranking and visualization, showing which items are agreed with the most across those divides---which it calls ``group aware consensus''. The allocation process here allocates the most prominent attention slots to items that have the most agreement across divides, highlighting those items and encouraging people to riff on them, suggesting variations that attract yet broader support \cite{small2021}.
    \end{exampleResponse} \vspace{-5mm}

    \begin{exampleFacilitation}[Attention allocator]
        Facilitators can promote or discourage certain kinds of attention allocation through the way they structure the deliberations---such as by giving certain people the floor at certain times---with goals (a qualitative value model) including the maintenance of baseline civility and ensuring the group delivers on its remit.
    \end{exampleFacilitation}

    Often, multiple attention allocators are required to meaningfully model a given situation. For example, consider a person browsing a social media feed. In this scenario, there are at least two attention allocators involved. The recommender system is an upstream attention allocator, algorithmically determining which of many potential atomic allocations (that is, positions of content within the feed) to realize. But the positioning of content within a recommender feed does not wholly determine what the person pays attention to. The person has agency too, and acts as their own, personal attention allocator deciding which of the many possible ways of allocating their continuous, first-person experience to ultimately enact or realize. We can say that the individual \newterm{partially delegates} their attention allocation to the recommender. More formally, this means that the recommender (as an attention allocator) influences the set of potential allocations that are available as inputs for the person's own, downstream attention allocator.\footnote{This phenomenon of attention allocators feeding into each other can be called an \newterm{attention stack} or \newterm{attention delegation network} depending on the structure.} 

    \begin{figure}[H]
        \centering
        \includegraphics[width=0.8\textwidth]{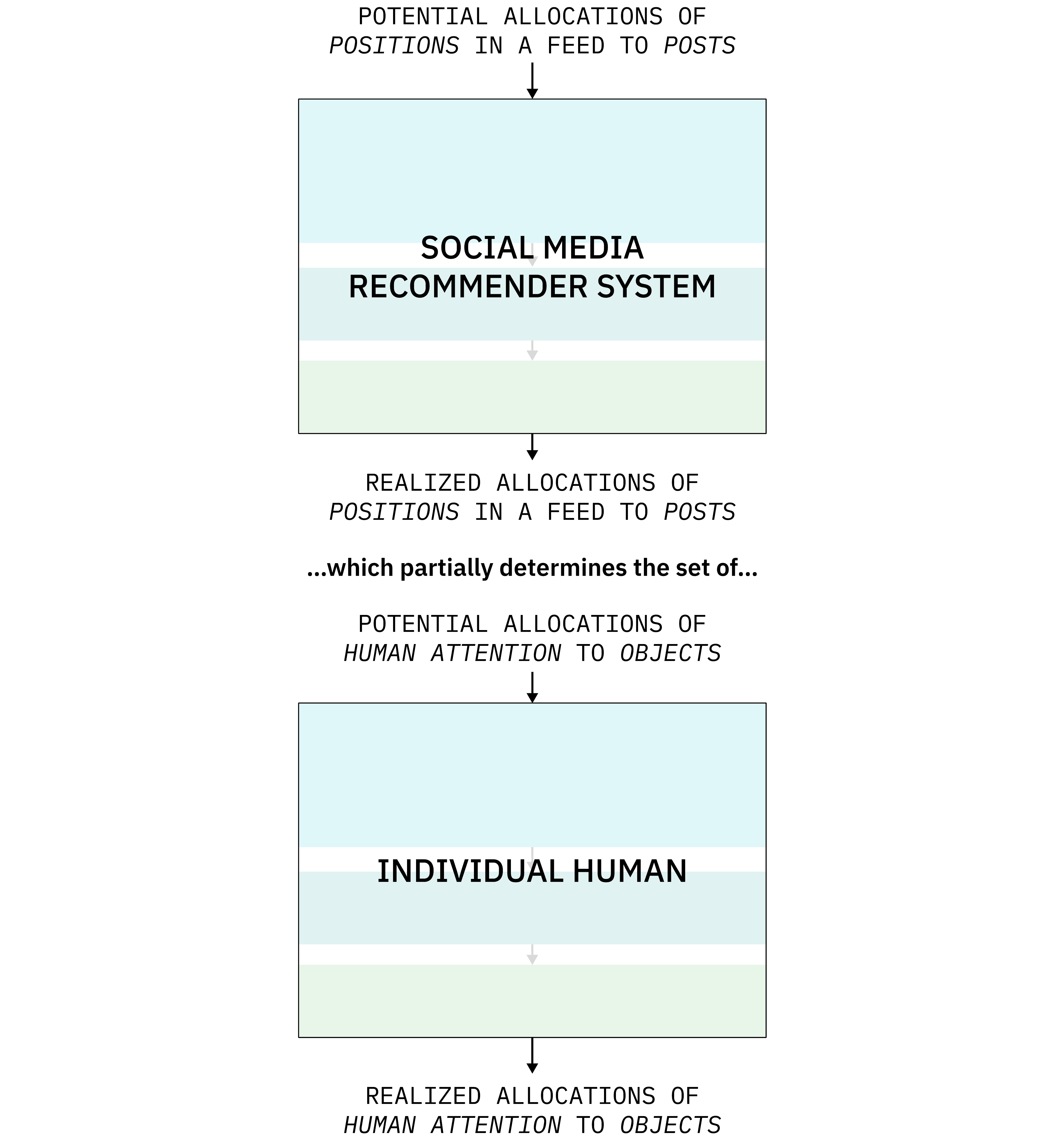}
        \vspace{2mm}
        \caption{\highlight{Example of two attention allocators chained together.} The allocation processes of both recommender systems and individual humans play a role in how human attention is allocated, because people partially delegate the allocation of their attention to the recommender. For simplicity, we have not included the learning processes of either the human or the recommender on this diagram.}
        \label{fig:stacked-allocators}
    \end{figure}

    While our focus here is human attention, under our definition, some algorithmic systems can also be said to attend to things. We chose a definition with this property because such systems can also significantly impact people depending on what information they process. 
    
    For example, consider a large language model (LLM) along with the infrastructure used to fine-tune that model. If a recommender system (in particular, the allocation process of a recommender system) is used to select content on which to fine-tune the LLM, then both the LLM and the recommender are attention allocators, with the LLM (in combination with its fine-tuning infrastructure) also ``attending'' to the recommender system. Because the recommender's allocation system is not updated during the allocation (as is standard), it is \textit{not} attending to the content it recommends. \footnote{However, separately, the recommender's learning system may be used to train the allocation system; such a learning process \textit{would} involve attending as it materially changes the system. There are some subtleties here, as our definition of attention is \textit{relative to the system boundary} that is being examined, which can include multiple people, algorithmic systems, organizations, etc. As another example, if someone uses an LLM to summarize an article, and then reads the summary, that summarizer is an attention allocator, but it is \textit{not} attending; since its state is not changing; it is simply outputting. However, the combined system of the person and the summarizer together \textit{are} attending to the article. Similarly, if a person attending to an article is an employee of an organization, that organization (system) can also be said to be paying attention to the article.} This becomes increasingly relevant with LLM-based systems being deployed to directly execute goals in the world.%

\subsection{The Optimization Stack}

\begin{figure}[H]
        \includegraphics[width=\textwidth]{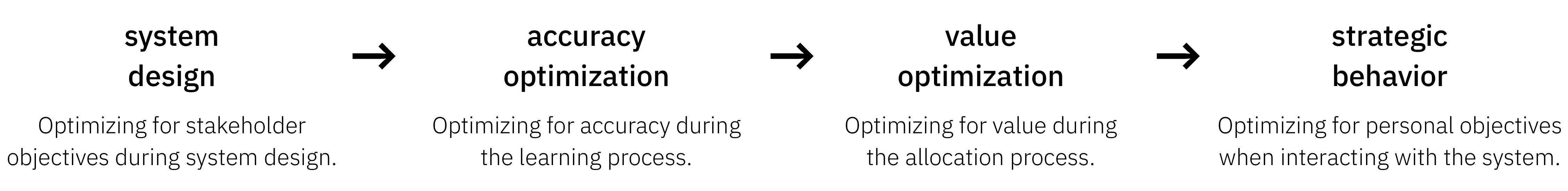}
        \vspace{-5mm}
        \caption{\highlight{The optimization stack.} The four levels of optimization that take place within (or adjacent to) an algorithmic attention allocator, such as a recommender system.}
        \label{fig:multilevel-optimization}
    \end{figure}

    To reason effectively about attention allocators (and allocators in general), it is important to recognize that they involve multiple levels of optimization (Figure \ref{fig:multilevel-optimization})---what we call an \newterm{optimization stack}.%

    The combination of \newterm{accuracy optimization}\footnote{Note that we can only observe phenomena like bridging-ness or entertaining-ness indirectly via measurable outcomes, such as the diversity of people who comment on a post or the watch time on a YouTube video. Thus, whenever we use the word ``accuracy'', we mean accuracy as measured against these measurable outcomes, which are proxies for the phenomena of interest. In many cases, there may not be a literal ground truth against which to measure accuracy, as it may not be possible to know the true impacts depending on what is predicted (e.g., we can't know true bridging-ness or entertaining-ness, we rely on proxies for them).} during the learning process and \newterm{value optimization} during the allocation process is an example of bilevel optimization \cite{colson2007}. But in algorithmic attention allocators (such as recommender systems), there are at least two other levels of optimization. Downstream of value optimization is the fact that individuals---both producers and consumers of (potential) allocations---will strategically optimize their behavior to further their own goals. Upstream of accuracy optimization are decisions made about the design of the learning and allocation processes. What impacts should be predicted? How should they be weighted in the value model? What data should be collected? Each of these questions will be answered to ``optimize'' some (perhaps qualitative) measure of the value of the system as a whole.\footnote{\emph{Reward Reports} provide an approach for understanding the components and interactions within this optimization stack \cite{gilbert2022}.}

    At all levels of optimization, quantitative \newterm{signals} and \newterm{metrics} are used to quantify the degree to which optimization efforts are successful. As signals and metrics are a significant focus of the remainder of this paper, we have included a list below which summarizes the most common ways in which they are used for optimization. In short, \newterm{signals} are used for allocation or ranking, and \newterm{metrics} are used for the evaluation of a system.

    \begin{enumerate}

        \item \highlight{Ranking.} If the value of a signal (or set of signals) is predictable or known for each of a set of alternatives, the alternatives can be directly ranked from most favorable to least favorable according to those signals, and the most favorable alternative(s) chosen. For example, value optimization may consist of evaluating the value model for each potential atomic allocation, and then promoting those allocations which are deemed most valuable.
    
        \item \highlight{A/B Testing.} If the value of the metric is not easily predictable, then experiments or A/B tests can be performed to produce estimates of the causal effects of each alternative intervention on the value of the metric. For example, system design often consists of conducting a large number of A/B tests to inform which design changes are implemented.
    
        \item \highlight{Machine Learning.} If the process being optimized is a machine learning model, then signals and metrics can be included as part of the loss or reward function on which that model is trained. For example, a reinforcement learning-based value optimization process could be trained to maximize the value (according to the value model) of the allocations it realizes.
    
    \end{enumerate}

    All of these have human facilitation analogs. A facilitator may directly rank topics to prioritize for discussion, may experiment with different methods of structured deliberation to see which work best in a given context, and will be constantly learning over time what facilitation strategies to use, according to their qualitative, internal ``metrics'' that measure the degree to which deliberation is successful.

\subsection{Bridging as a Property of Attention Allocators}
\label{sec:bridging-property}
    The \newterm{bridging goal} is an increase in mutual understanding and trust across divides, creating space for productive conflict, deliberation, or cooperation.\footnote{In principle, one way to promote bridging would be to cause people to value bridging impacts more within their own, personal attention allocators. However, having people unilaterally change their behavior at a significant scale is a difficult collective action challenge and faces significant psychological friction; thus our focus is on algorithmic and institutional bridging systems.} We say that an attention allocator is \newterm{bridging} to the extent that it causally supports the bridging goal through its allocations. 
    
    Bridging is thus a system property, rather than a distinct kind of system, and falls on a spectrum---attention allocators may either facilitate or discourage such allocations to varying degrees (and are thus more or less bridging). For example, a recommender system that shows people content that leads to an overall increase in mutual understanding and trust could be considered bridging. If a recommender system has a parameter that determines how much to weigh ``bridging signals'' versus ``pure engagement'' signals, then depending on the setting of that parameter, the ``same'' recommender system might be significantly bridging or extremely polarizing. 
    
    While this qualitative definition of bridging is not directly operationalizable, in the remainder of this paper, we provide formalisms that make an approximation of bridging actionable using models, signals, and metrics. In particular, there are several places in which bridging can be incorporated into attention allocators, at different parts of the optimization stack, some of which are indicated by bridge icons in Figure \ref{fig:allocation-process} and \ref{fig:learning-process}. For example, the system can model relations present in a population and can implicitly predict, in the \newterm{impact prediction} stage of the allocation process, whether divisions will increase or decrease. These predictions can then be included as signals in the value model and influence which allocations of attention are realized \cite{stray2021b}. At the level of system design, \newterm{bridging metrics} could be considered and used to adjudicate which of a set of potential interventions should be implemented. In Sections \ref{sec:data-and-modeling}-\ref{sec:metrics}, we give concrete examples of such signals and metrics, and describe how attention allocators---particularly those that are algorithmic---can be designed to be bridging.

\clearpage

\section{Data and Modeling}
\label{sec:data-and-modeling}

    Above, we stated that an attention allocator is bridging to the extent that it supports the ``bridging goal'': increasing mutual understanding and trust across divides, creating space for productive conflict, deliberation, or cooperation. This qualitative property must be formalized if we are to extend the insights from offline bridging practices that have been developed over millennia into digital attention allocators such as recommender systems and collective response systems. %

    In the next three sections, we describe how the notion of bridging can be formally represented and then quantified. In particular, we: introduce the concept of a \newterm{relation model}, a representation of the relationships or affinities between individuals in a given population (Section \ref{sec:data-and-modeling}); describe signals which can be used as the basis for ranking and allocation (Section \ref{sec:signals}); and describe metrics which can be used to measure the impact of an entire attention allocation system (Section \ref{sec:metrics}).

\subsection{Data}

    To model bridging, you need relevant data. Different systems use different approaches for collecting or eliciting such information, and the data can take different forms. Examples are given below from our three example domains.

    \begin{exampleRecommender}[Data Elicitation]
        Users of online platforms generate engagement data such as likes, reactions, comments, shares, dwell time, tagging, direct messages, and so on. Recommender systems use this data to learn about the preferences and perspectives of users. Implicitly, this data reveals how users differ, and the degree of affinity between them.
    \end{exampleRecommender} \vspace{-5mm}

    \begin{exampleResponse}[Data Elicitation]
         Each instance of a collective response system is focused around an issue, prompt or question. Participants can either provide new responses or vote on existing ones. The user interfaces generally differ from those of conventional forums such as Reddit. For example in Polis, people are shown a succession of responses one by one, and can choose to agree, disagree, or pass in response to each statement. Polis' algorithm selects which statement to show next using a number of criteria, one of which is to maximize learning about the relations between participants, responses and, implicitly, each other~\cite{small2021}.
    \end{exampleResponse} \vspace{-5mm}

    \begin{exampleFacilitation}[Data Elicitation]
        Facilitators ``reading a room'' identify a host of subtle cues over the course of deliberation. Common examples include flared nostrils, changes in breathing, and hushed silences. Structured exercises may also be used to sort the room into groups of perspectives \cite{carson2020}.
    \end{exampleFacilitation}

    This data contains information about people's preferences, perspectives, opinions, identities, or worldviews. Implicitly, it provides information about the relationships, affiliations, and affinities that exist in a population. In many cases, this ``data'' would have been ``collected'' anyway. For example, an effective facilitator will intuitively gather such information in the course of interacting with a group. Similarly, a recommender system where users interact with each other will implicitly be eliciting data useful for modeling human relationships. The data elicited by such systems is often very contextual---it is highly dependent on system/process design, affordances, culture, the environment, existing divisiveness, and many other factors.

    \begin{questions}[Eliciting Data for Bridging]
        \item What affordances provide the most useful information about relationships in a population? (Taking into account that respect for privacy is valuable both ethically and for gathering accurate data.)
        \item Can elicitation methods from one domain (say, mini-publics) be translated to another (say, recommender systems)?
        \item How can we mitigate and account for the fact that the act of eliciting information about relationships can itself influence those relationships?
    \end{questions}

\subsection{Relation Model}

    \noindent
    A \newterm{relation model} is a formal representation of the relationships between people in a population, which can be learned or inferred based on the data available. These relationships could be explicit (e.g., friendships) or implied (e.g., the affinities between people who have similar preferences or worldviews). The relation model is a ``state model'', describing a snapshot of the world at a given point in time.

    Formally, a relation model---as we define it---can be decomposed into three components: \variable{people} (the people who interact with the system), \variable{items} (the alternative objects to which \variable{people} may attend, which may also be people), and \variable{relations} (the one-to-one relations between \variable{people} and \variable{items}, intended to capture goodwill, agreement, affinity, reactions, or similar). This general framework highlights a common structure shared by our three examples, as summarised in Table \ref{table:components}.

    \note{Kinds of relation model: people \(\to\) beliefs; people \(\to\) beliefs about what other people believe (perception gap); people \(\to\) people (communication, e.g., followers); people \(\to\) people (reputation); people \(\to\) content (e.g., likes)}

    \begin{table}[H]
        \sffamily
        \begin{tabular}{@{}lp{4.1cm}p{4.9cm}p{3.8cm}@{}}
         & \faSort \textbf{Recommender System} & \faClipboard \textbf{Collective Response System} & \faComments \textbf{Human Facilitation} \\ \addlinespace
        \variable{people} & users & participants & participants \\ \addlinespace
        \variable{items} & posts / content & comments & claims / positions \\ \addlinespace
        \variable{relations} & revealed preferences & agree, disagree, or pass & qualitative opinions \\
        \end{tabular}
        \caption{Recommender systems, collective response systems, and human facilitation all share a common structure.}
        \label{table:components}
    \end{table}

    The three components will be modeled differently in different contexts. Most comprehensively, the relation model may simply be the totality of all available data available about the population. More structured models can also be used, two common examples of which are \newterm{graph-based} models and \newterm{space-based} models.

\begin{figure}[t]
        \includegraphics[width=\textwidth]{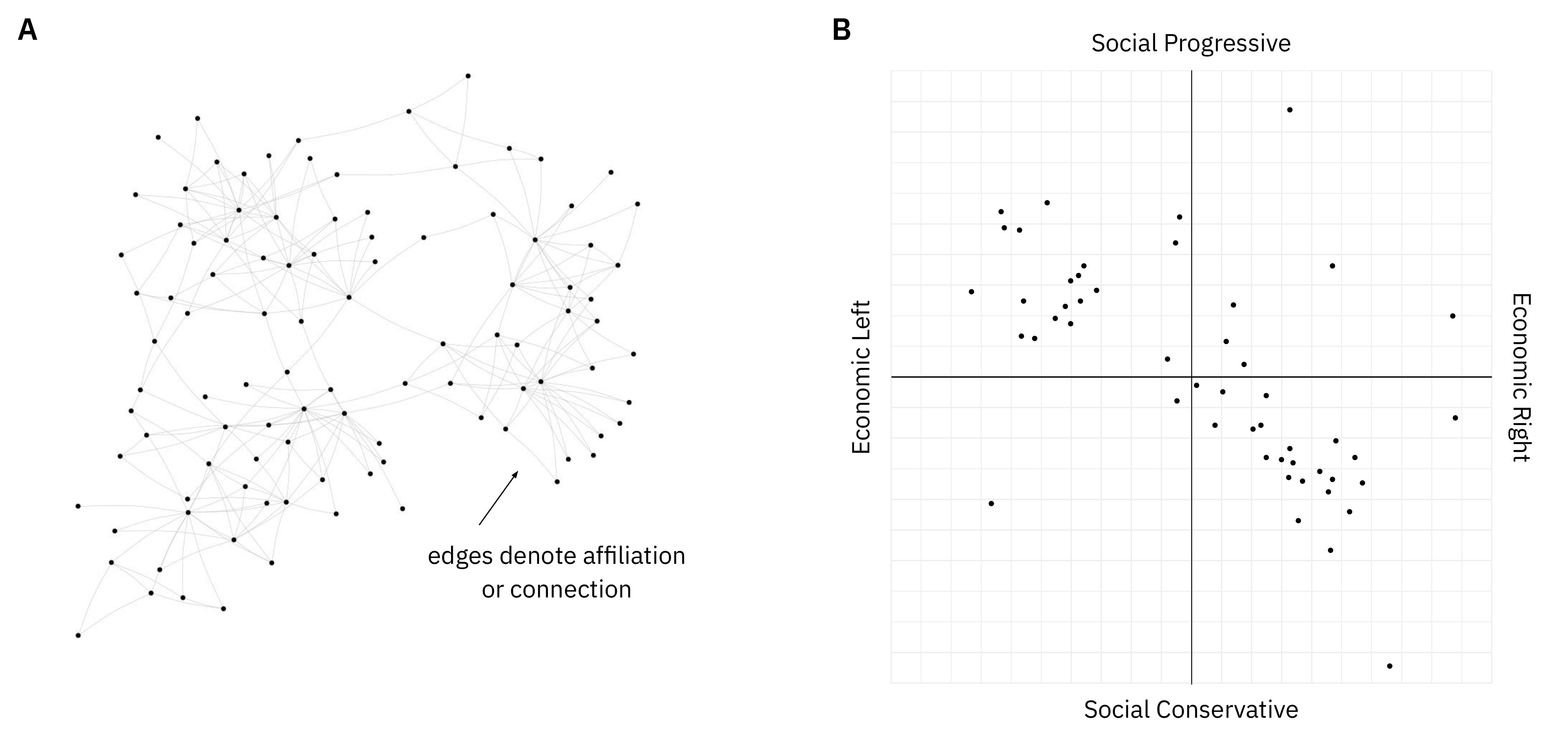}
        \vspace{-5mm}
        \caption{Simple examples of (a) a graph-based model and (b) a space-based model.}
        \label{fig:model-types}
    \end{figure}

    \paragraph{Graph-based} Graph-based models represent \variable{people} as nodes in a (mathematical) graph or network. The edges in the graph characterize the relationship between people, and can represent explicit, active communication channels, or be weighted to represent more fine-grained and abstract types of affiliation. A simple example of a graph-based model is given in Figure \ref{fig:model-types}(a).

    \paragraph{Space-based} Space-based models represent \variable{people} as locations within an ambient ``opinion space''. The similarity between people's opinions, preferences or viewpoints is characterized by how close to one another they are in this space. For example, people may be modeled as a location on the left-right political spectrum (a one-dimensional space), or assigned a position on a political compass (a two-dimensional space). A simple example of a space-based model is given in Figure \ref{fig:model-types}(b). In general, the space might have hundreds or thousands of dimensions. The \term{items} may also be represented as points in the same space, in which case the proximity of a person to an item represents the degree of goodwill, agreement, or other ``favorable'' relationship between them.\\ 

    These two approaches to relation modeling are not exhaustive. For example, it may make sense to model relations at a higher level of abstraction, because it is not necessary to model individuals to represent useful information about divisions in a population. A very simple relation model might be: \{70\% of people like \emph{The Beatles}, 50\% like \emph{Adele}, 10\% like \emph{Nickelback}\}\footnote{We use musical artists as an example, but these could equivalently be levels of support for different political positions. For example, Facebook's identification of news sources that are seen as broadly trustworthy might be considered an example of such a model \cite{2018HelpingEnsureNews}.}. This tells us that there is overlap between people who like The Beatles and Adele, that both are fairly popular, and that Nickelback is comparatively not. Such \emph{aggregate models} could incorporate item classifications (e.g., Nickelback might be labeled as non-bridging), or be broken down by subgroups. Note, however, that such models are likely less expressive than graph or space-based models, and depending on the percentages it will not always be possible to infer overlap.

    \begin{exampleRecommender}[Relation Model]
        The recommender systems on modern social media platforms are often built using deep neural networks that learn a numerical representation of each user and item. These vectors or ``embeddings'' are usually high-dimensional (hundreds or thousands of dimensions), and correspond to a position within a latent embedding space that characterizes each user's history of behavior on the platform and, by implication, their opinions, preferences, and worldview. Thus, these embeddings constitute a space-based model \cite{covington2016,satuluri2020}.

        \quad Social networks also often have an underlying graph-based structure, such as graphs of friends on Facebook, or follow networks on Twitter. Such networks constitute graph-based models of \variable{people}. There are other possible graph structures. For example, you could consider a single graph where both \variable{people} and \variable{items} are vertices, and edges are used to indicate interactions between \variable{people} and \variable{items}. In some cases, information from such graph-based models is translated to space-based models for use in recommender systems \cite{satuluri2020}.
    \end{exampleRecommender} \vspace{-5mm}
    
\begin{figure}[t]
        \includegraphics[width=\textwidth]{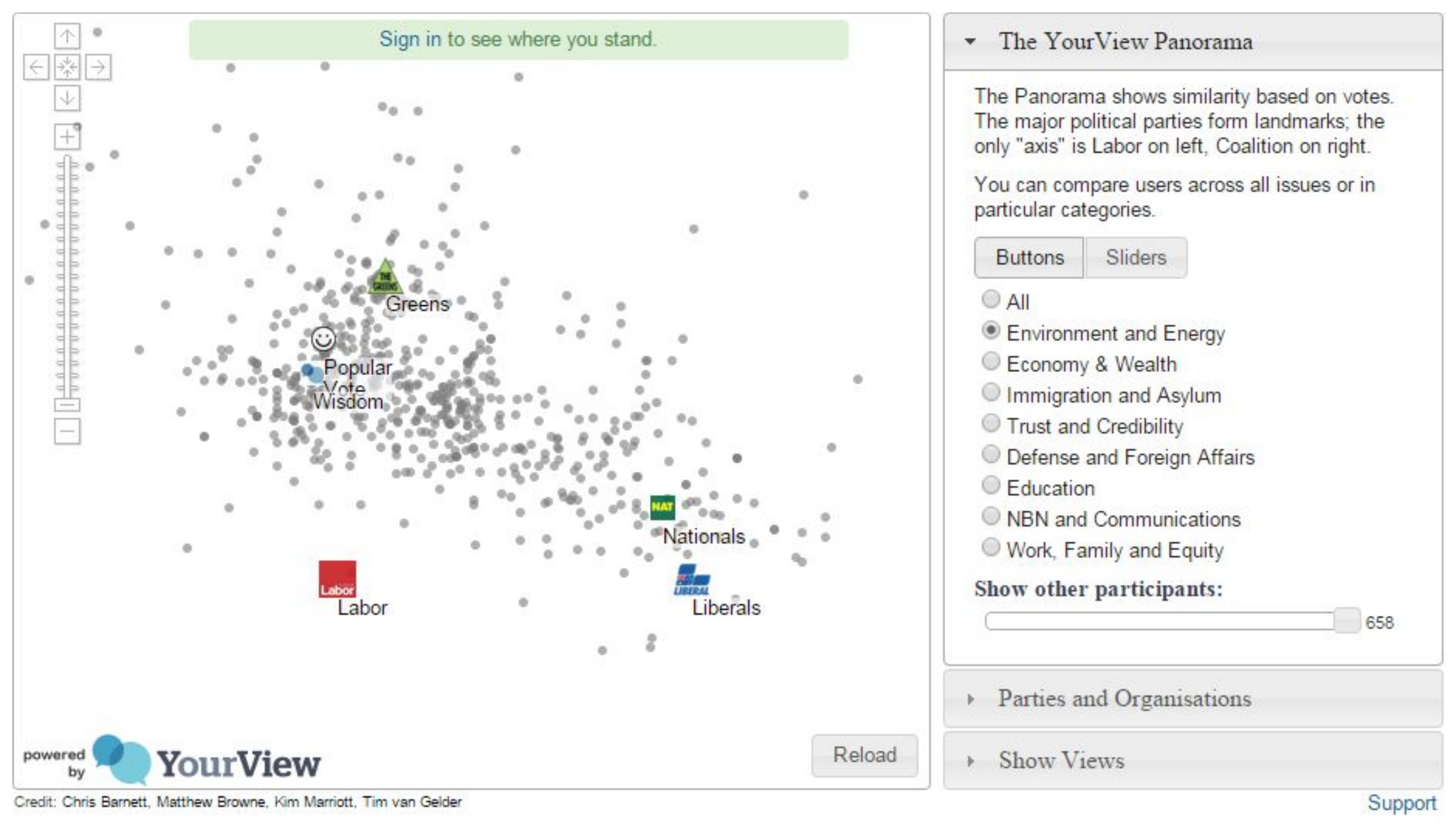}
        \vspace{-6mm}
        \caption{The YourView ``Panorama''---a two-dimensional space-based model.}
        \label{fig:panorama}
    \end{figure}

    \begin{exampleResponse}[Relation Model]
        
        Collective response systems like YourView or Polis involve \variable{people} submitting \variable{items} (called comments or responses), and voting on items shown via a recommendation algorithm. These votes are represented by a matrix where rows correspond to \term{people}, columns correspond to \term{items}, and the cell values indicate votes: ``Agree'' (encoded as $+1$), ``Disagree'' ($-1$), or ``Pass'' ($0$). The rows in the vote matrix can be viewed as the locations of people in a space-based model. So that it can be visualized, both YourView and Polis compress this relatively high-dimensional representation into a two-dimensional space-based model, positioning people closer together if their votes are more similar. Intuitively, this means that if two people voted the same way on every item, they will end up at the same point \cite{small2021}. An example of the YourView space-based model is shown in Figure \ref{fig:panorama}.
    \end{exampleResponse} \vspace{-5mm}
    
    \begin{exampleFacilitation}[Relation Model]
        Facilitators pay attention to how people relate to each other, and where they fall along the salient axes of disagreement within the groups that they are facilitating, thus intuitively applying qualitative versions of both graph-based and space-based models.
    \end{exampleFacilitation}
    
    By introducing the term ``relation model'' we are not aiming to prescribe a particular kind of model or claiming to have invented one, but merely aiming to describe a \emph{role} that certain models can play within an attention allocator. Models from many existing fields (social choice, opinion dynamics, recommender systems, etc.) could be used as a relation model. The new term also helps when talking about the commonalities across recommender systems, collective response systems like YourView and Polis, and human-facilitated deliberations.

\begin{figure}[!ht]
        \includegraphics[width=\textwidth]{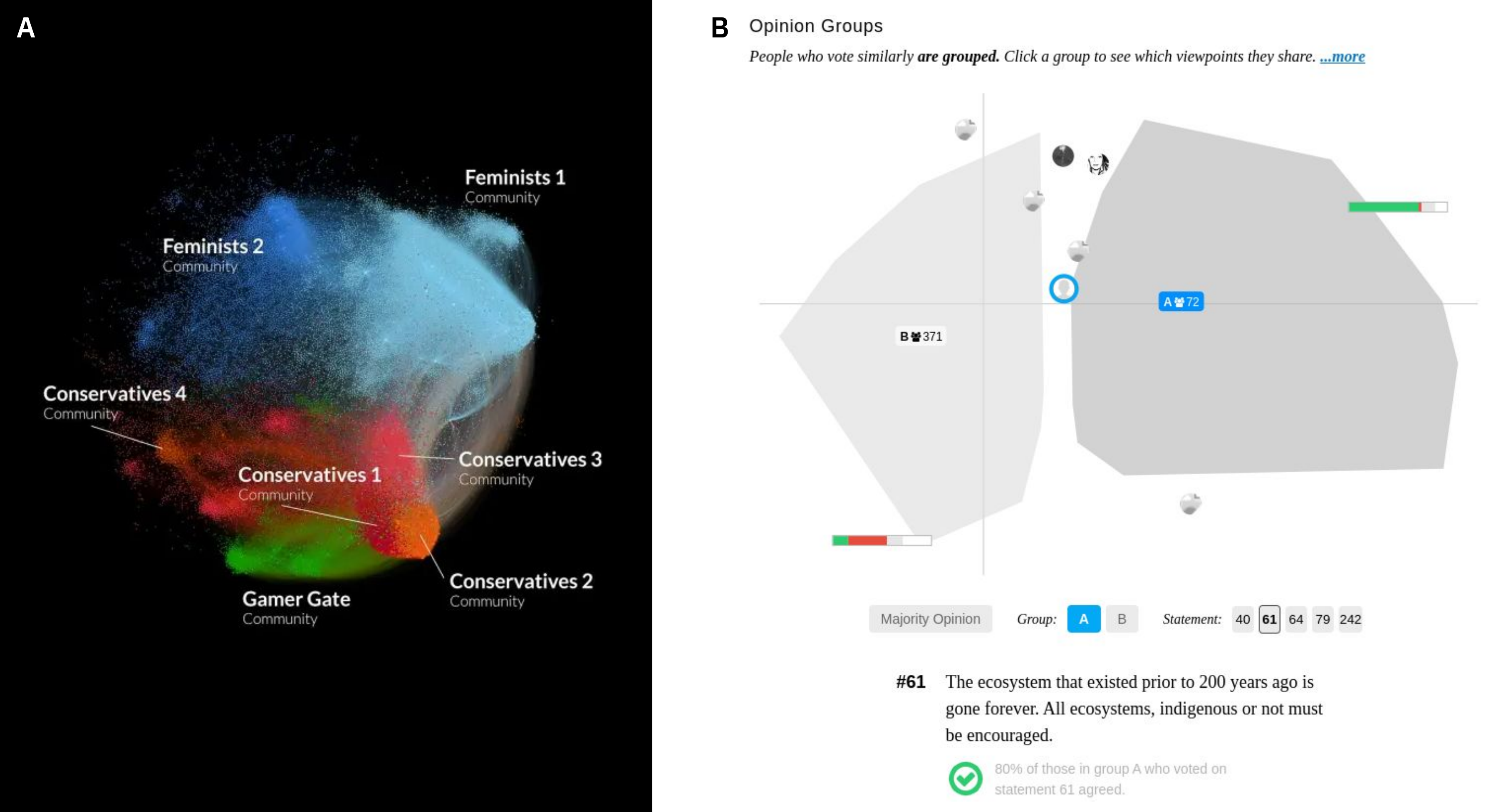}
        \caption{\highlight{Examples of clustering in relation models.} Figure (a) is a graph-based model of Twitter users, clustered by their stance regarding the legal status of abortion. Figure (b) is a space-based model generated by Polis, where a New Zealand newspaper elicited a collective response on the topic of protecting biodiversity. \textit{Image credits: \textcite{clifton2015} and \textcite{scoop2019}.}}
        \label{fig:clustering}
    \end{figure}

    How can information about the quality of relations be extracted from space- or graph-based models? In both cases, the models can be used to identify clusters or groups of people who think similarly. Divisions can be thought of as the spaces between these groups: how far apart they are, how they think about each other, and how they interact. Annotating the models with such additional structure can provide insight into the nature and strength of divisions that exist in a population. Figure \ref{fig:clustering} presents examples of this sort of clustering in both graph-based and space-based models.

    \begin{exampleResponse}[Clustering]
        In Polis, \variable{people} are then clustered into two to five distinct groups, an example of which is shown in Figure \ref{fig:clustering}(b). Clustering is performed directly on the two-dimensional projection using the $k$-means clustering algorithm, and a goodness-of-fit statistic is used to determine which number of clusters best fits the vote data \cite{small2021}.
    \end{exampleResponse}

    \begin{questions}[Relation Models]
        \item What kinds of representations are appropriate for a given context? For example, discrete clusters (Polis) versus continuums (Twitter Community Notes).
        
        \item How can these representations be best operationalized in that context? For example, Polis currently uses PCA to project the vote space into two dimensions, sets $k$ for $k$-means clustering to 2-5 groups, and chooses the best $k$ using silhouette coefficient to identify clusters \cite{compdem2022}.

        \item How to model and identify divides relevant for bridging (e.g. deep political divides), and distinguish them from benign divides between certain communities or interest groups (e.g., ``librarians'' and ``people who like fishing'').
        
        \item Are the data collection processes and relation models of existing attention allocators, such as recommender systems, sufficient to perform bridging-based ranking?
    \end{questions}

\clearpage

\section{Signals}
\label{sec:signals}

    Signals are numbers that are aggregated into an overall measure of value and used as the basis for value optimization during an allocation process.\footnote{We use the term ``signals'' because it is used by platforms who manage recommender systems (e.g., \cite{meta2023}). In our more general model of an attention allocator (Section \ref{sec:attention allocation-systems}), signals correspond to the predicted impacts.}

    To make attention allocators more bridging, we need signals which indicate the degree to which a potential allocation is likely to be bridging. For example, we want to be able to quantify the degree to which recommending a particular Facebook post, or presenting a particular claim in Polis, is likely to achieve the bridging goal.

    To aid discussion, we draw two distinctions between different kinds of signals. The first distinction is that signals can be \newterm{observed} or \newterm{predicted}, depending on whether the information they represent is already known.\footnote{When predicted, they can be included among the predicted impacts of an allocation in the allocation process of an attention allocator (Figure \ref{fig:allocation-process}), and included in the value model according to which allocations are selected. When observed (retrospectively), they can be used for monitoring, and compiled into datasets to train the models used for impact prediction.} The second distinction is that signals can be \newterm{causal} or \newterm{heuristic}, depending on whether allocating attention to the corresponding object is known to causally contribute to bridging, or is merely thought to do so.\footnote{We expect that in practice most signals will be heuristic due to the impracticality of inferring, at scale, the causal impact of every individual item of content. However, such causal inference has been done in small settings \cite{voelkel2022}.}

\subsection{Practical Examples}
\label{sec:signal-examples}

    In this section, we describe three approaches to constructing signals for bridging systems and give concrete examples of each.

\subsubsection{Motifs}

\begin{figure}[!ht]
        \centering                                                                               \includegraphics[width=0.9\textwidth]{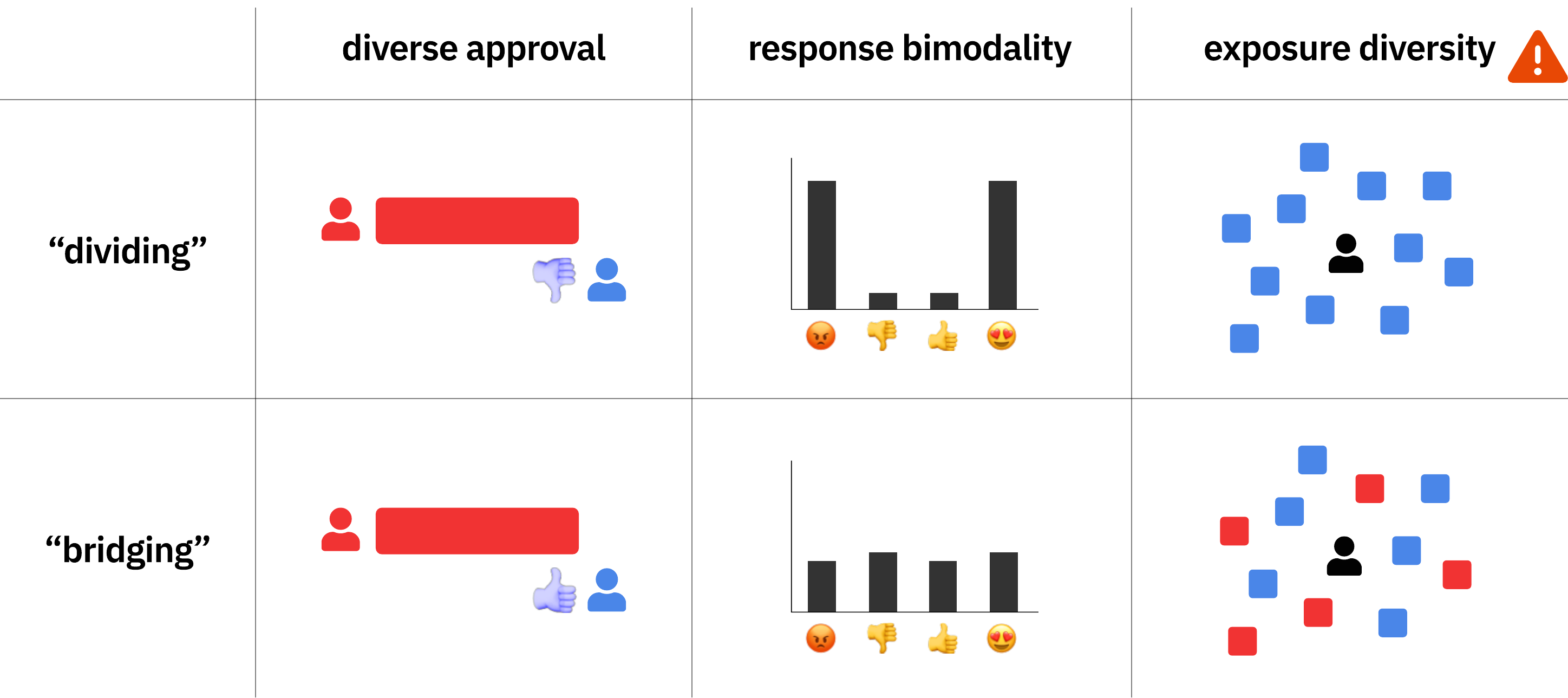}
        \caption{\highlight{Examples of motifs that could form bridging signals.} Note that evidence for the validity of all three motifs is limited, and in particular, some evidence suggests that use of the exposure diversity motif (\faWarning) may \textit{increase} affective polarization (see main text).}
        \label{fig:motifs}
    \end{figure}

    Signals can be built on \textit{motifs}, which are patterns of interaction thought to be associated with certain outcomes---in our case, bridging. Here we describe three previously-proposed motifs: diverse approval, response bimodality, and exposure diversity. Visual intuition for each of these motifs is given in Figure \ref{fig:motifs}.

    \paragraph{Diverse approval} The first motif, \newterm{diverse approval}, states that bridging occurs when a person creates an item that is approved of, supported, or otherwise validated by people who would normally disagree with them.\footnote{It is theoretically straightforward to determine when there is a divide between two people; this can be quantified by a metric between their respective vertices or positions in graph- or space-based relation models. Such divides are already inferred automatically by many existing attention allocators.} Such ``surprising validation'' \cite{sunstein2012,reisman2012} is thought to correlate with accuracy and good faith communication, while also prompting observers to consider more perspectives (e.g., ``if someone like that disagrees with me, maybe I had better rethink'' \cite{sunstein2012}). A slightly stronger version of the heuristic states that bridging occurs when an item is endorsed by people from multiple (i.e., more than two) diverse viewpoints. Versions of the diverse approval heuristic have been operationalized with success in at least two bridging systems.

    \begin{exampleResponse}[Outgroup Approval]
        In YourView, each participant had a ``credibility'' rating---a type of reputation score that quantified, primarily, the degree to which the comments they contributed on any particular issue attracted support from people who disagreed with them about that issue. Thus grounded, the definition of credibility could be extended recursively: YourView assumed that people were credible if they were respected or trusted by credible people they disagreed with. This circular definition is akin to the concept of ``eigentrust'' \cite{aaronson2014} or the logic of the PageRank algorithm, which has been applied in the context of social media to measure different forms of reputation \cite{allen2019,allen2022,reisman2003,reisman2018}.
    \end{exampleResponse} \vspace{-5mm}

    \begin{exampleRecommender}[Diverse Approval]
        Community Notes (formerly known as Birdwatch) is a Twitter feature that allows users to ``collaboratively add helpful notes to Tweets that might be misleading'' \cite{wojcik2022}. In its current form, Community Notes contributors can also rate notes contributed by others as ``helpful'', ``somewhat helpful'' or ``not helpful''. Notes are awarded a high helpfulness score (and hence displayed publicly) only if they have been ``rated helpful by raters with a diversity of viewpoints''. This is achieved using a version of the matrix factorization algorithm that is commonly used in recommender systems \cite{twitter2022a}.
    \end{exampleRecommender}

    \paragraph{Response bimodality} Under the second motif, \newterm{response bimodality}, attention allocations are said to be (relatively) bridging if the distribution of ratings or reactions elicited by the relevant item is not polarized. Intuitively, a distribution is most polarized if it is markedly bimodal (that is, it has a ``U'' shape). One line of research on the use of recommender systems for depolarization relies on this heuristic \cite{badami2017,badami2018,badami2021}. They quantify the degree of polarization in a rating distribution by training a binary classifier that takes in features computed from a histogram of item ratings and outputs whether or not the distribution is polarized, as trained on ``human expert'' classifications.

    \paragraph{Exposure diversity} The third motif, \newterm{exposure diversity}, states that bridging occurs when people attend to items from diverse sources, particularly from sources they don't normally see. This motif implies that to facilitate bridging, people should be shown items from outside their ``filter bubble'' or ``echo chamber''. However, a number of studies have failed to find evidence of algorithmic filter bubbles \cite{bruns2019}, and other research suggests that indiscriminately showing people posts from their outgroup may cause relations to deteriorate \cite{bail2018}. For these reasons, the validity of the exposure diversity heuristic is questionable.
    
    \vspace{5mm} \noindent
    The interaction patterns captured by the above three motifs are conceptually simple.\footnote{Though in most cases, their computational complexity remains an open question.} We can also define more complex and subtle motifs that, for example, model longer patterns of interaction, or incorporate more nuanced information about the people and items involved. These could be specified based on insights from bridging in offline settings \cite{powell2021,shigeoka2020}, or learned algorithmically. Below, we give an example of a learned class of motifs, \newterm{positive interactions across divides}, which generalizes the idea of ``diverse approval''.

    \begin{exampleRecommender}[Positive Interactions Across Divides]
        Consider a hypothetical social media platform named BridgeTok. BridgeTok built a dataset of posts with their subsequent social exchanges (reactions, comments, shares, etc.), along with human ratings of the degree to which each exchange was ``positive''. They then trained a model to predict the positivity of any arbitrary interaction pattern. These predicted positivity scores, combined with existing user embeddings, are used to quantify the degree to which new interactions represent ``positive interactions across divides''. The most bridging allocations of attention, according to this heuristic, are then selected for in the BridgeTok recommender system.
        
        \quad Note there is a precedent for this kind of modeling at Facebook: among the Facebook Papers provided by Frances Haugen there mention of experiments using ``diverse positive motifs'' to identify positive ``conversations'' and civic interactions \cite{facebook2022b,facebook2022,facebook2022a}.
    \end{exampleRecommender}

    A focus on patterns of interaction, without consideration of the meaning of natural language content, increases the extent to which motif-based signals can be implemented internationally, across languages. That said, the effectiveness of a given ``structural'' heuristic for identifying bridging behaviors may depend on cultural and social context, and in some settings content-based signals (Section \ref{sec:content}) may be useful or necessary.

\subsubsection{Surveys}

\begin{figure}[!ht]
        \centering                                                                               \includegraphics[width=\textwidth]{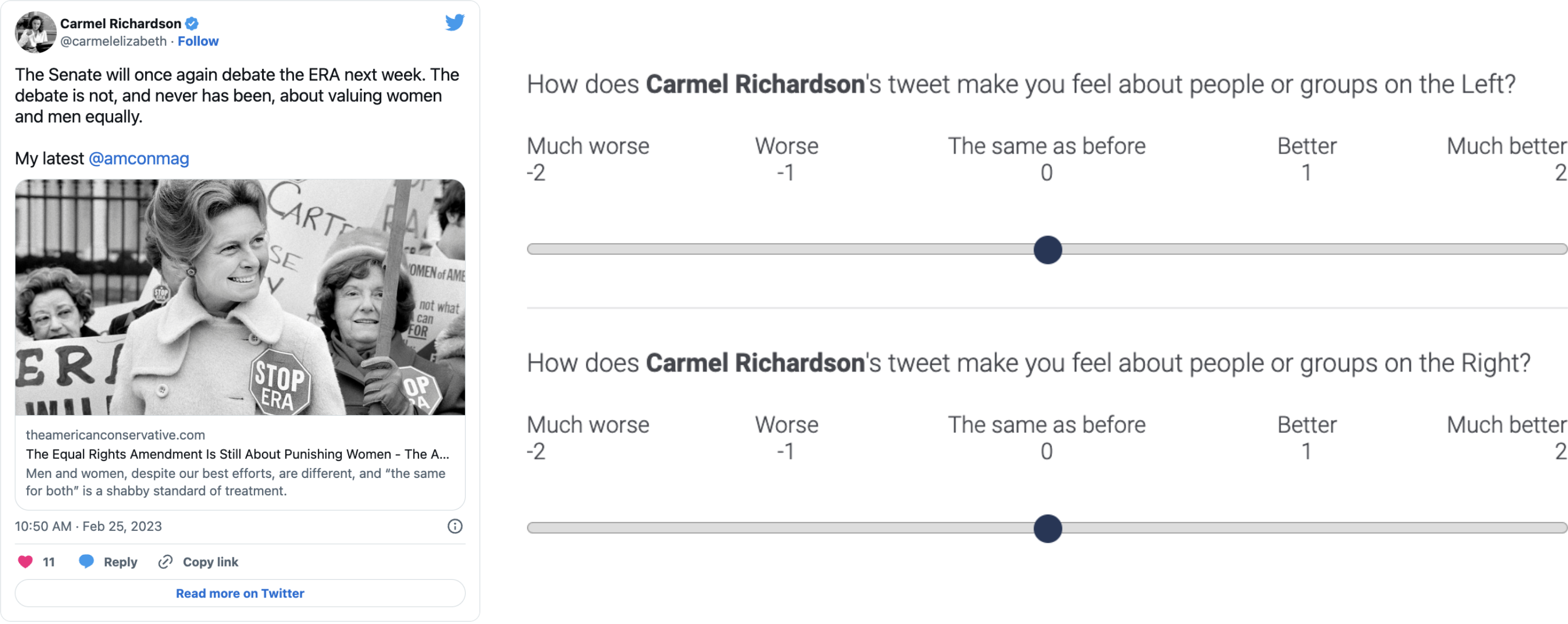}
        \vspace{-5mm}
        \caption{\highlight{Example of a survey that could form a bridging signal.} This example is taken from \textcite{milli2023}. }
        \label{fig:surveys}
    \end{figure}

    Signals can also be built on survey questions, the responses to which indicate the degree to which a given allocation of attention is bridging. An example of such survey questions is given in Figure \ref{fig:surveys}. 

    While surveys are burdensome for users to complete, it is not necessary to ask users to respond to a survey about everything they see. Instead, each user's response can be \textit{predicted} using a model trained on a much sparser dataset, and these predictions used as signals.

    There is a huge number of potentially relevant questions that could be asked. A useful starting point for possible questions is Stanford's \textit{Strengthening Democracy Challenge} \cite{voelkel2022}, which included many relevant survey questions as outcome variables, including questions about partisan animosity, support for partisan violence, support for undemocratic practices, opposition to bipartisan cooperation, social distrust, social distance, anger towards outpartisans, empathy with outpartisans, perceived threat of outpartisans, and strength of partisan identity. In particular, affective polarization or partisan animosity is commonly measured using a \newterm{feeling thermometer}.

    \paragraph{Feeling thermometer} The feeling thermometer is a short survey instrument commonly used to measure affective polarization. In its simplest form, it consists of asking people to rate their feelings towards their country's two largest political parties on a scale from 0 (cold) to 100 (warm). The difference between these two scores indicates the degree of affective polarization. \textcite{stray2021b} proposed incorporating this short survey instrument into social media platforms. Survey responses could be linked with items recommended to the user shortly prior to their completing the survey. In this way, the degree to which responses trend towards less affective polarization could be used to quantify the degree to which recently shown items of content were bridging.%

\subsubsection{Content}
\label{sec:content}

\begin{figure}[!ht]
        \centering                                                                               \includegraphics[width=0.9\textwidth]{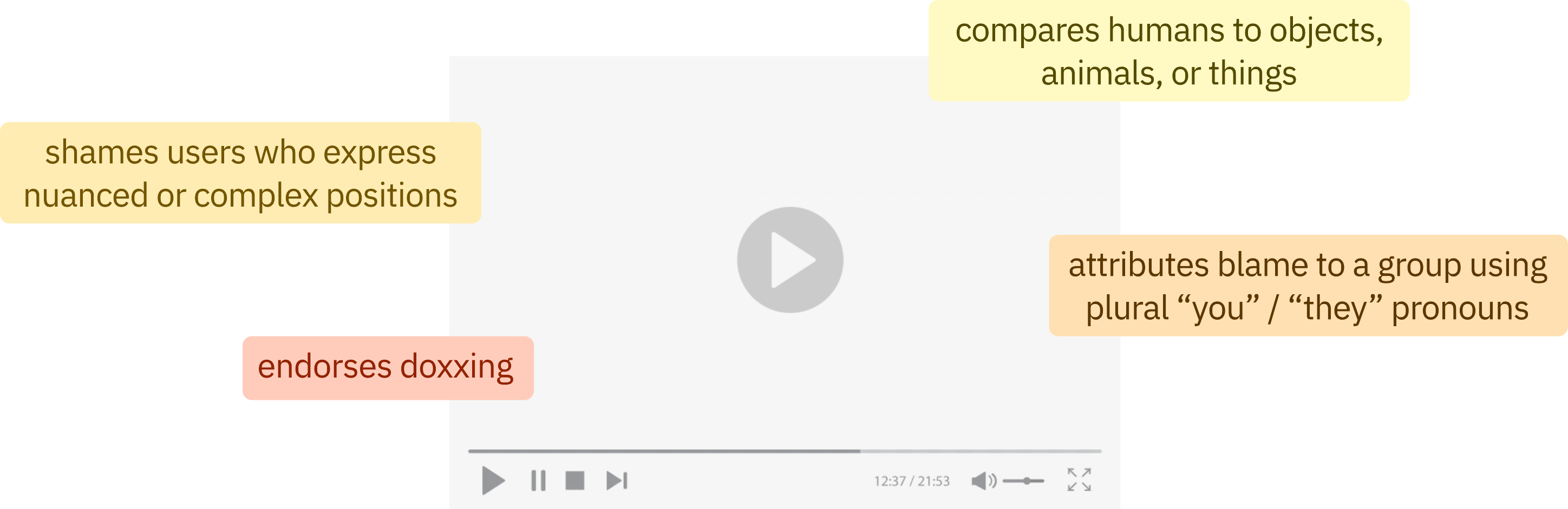}
        \vspace{2mm}
        \caption{\highlight{Examples of content labels that could form a bridging signal.} These example are taken from the ``archetypes of polarization on social media'' framework described in \textcite{hawke2022}.}
        \label{fig:content}
    \end{figure}

    Finally, signals can be based on (semantic) properties of the content itself. For example, one can use a machine learning model to estimate the degree to which content is dehumanizing or contemptuous. More examples of more complex properties relevant to bridging are given in Figure \ref{fig:content}.

    As with surveys, there is a large number of semantic properties of content that are potentially relevant for bridging. A useful starting point for possible properties to target is the ``archetypes of polarization on social media'' framework described in \textcite{hawke2022}. In this framework, polarization is decomposed into five archetypes---attitudes, affiliation, interaction, interests, and norms---and for each of these archetypes there are several examples given of concrete online behaviors that reflect that archetype, most of which depend on an understanding of the semantics of an item of content. For example, polarization of attitudes can be reflected by ``negative reactions (laughing, anger, etc.) to content with a group signifier'', ``keywords and hashtags that refer to the outgroup'', ``use of plural `you' and `they' pronouns in reply threads'', ``generalized blame and attribution'', and ``interaction with bias-affirming content''. Automated classifiers could be used to identify these kinds of semantic properties in items of content, and these semantic labels could be used as signals to prioritize allocations of attention that are bridging (or conversely, deprioritize allocations of attention that are divisive).

    \begin{questions}[Signals]
        
        \item How can we evaluate the quality of bridging signals? What are their relative strengths and weaknesses? Are there particular domains where one signal is more appropriate than another?
    
        \item How common are examples of diverse approval in practice (e.g., on modern social media platforms)? Are there enough examples to make a meaningful impact if they are featured more prominently by a recommender system?
        
        \item To what extent does giving bridging signals more weight in the value model actually incentivize bridging behavior? In economic terms, what is the ``value model elasticity of bridging''?

        \item How computationally complex are different bridging signals?

        \item To what extent are particular motifs valid across international contexts?

        \item How could bridging signals be gamed or abused? What would ``bridging-bait'' content look like?

        \item Some promising signals, such as diverse approval, may have significant bridging benefits but also incentivize shallow content (e.g., cat memes). How might they be refined or complemented by additional signals to incentivize deeper content? 
        
        \item What signals are useful for identifying where bridging is even appropriate at all (and how heavily to weight bridging signals), across different contexts, cultures, or types of content?

        \item How can the best bridging signals be effectively used in more federated or decentralized systems? If there are challenges such as deployments, under what circumstances are they fundamentally intractable?
        
    \end{questions}

\clearpage

\section{Metrics}
\label{sec:metrics}

    Metrics provide quantitative assessments of attention allocation systems; they are used both for ongoing monitoring and guiding system development (e.g., via A/B tests).

    To be useful, metrics need a clear normative interpretation. Within a given context, we should be able to say that higher metrics reflect success at achieving the bridging goal, or that certain configurations of multiple metrics are better than others.

    We distinguish between two kinds of metrics: \newterm{relation metrics} and \newterm{bridging metrics}. Relation metrics capture the quality or ``health'' of relations at a given point in time (as represented by the relation model). In contrast, bridging metrics capture the degree to which attention allocators are associated with an improvement or deterioration in relations, over time. To use a mathematical analogy, bridging metrics can be thought of as the ``derivatives'' of relation metrics with respect to time.

\subsection{Practical Examples}

    In this section, we describe two approaches to constructing metrics for bridging systems, and give concrete examples of each.

\subsubsection{Prevalence of Signals}

    Most simply, metrics can be built on top of signals (Section \ref{sec:signals}) that indicate good or improving relations in a population. For example, a relation metric could be defined to be the overall prevalence in which a positive motif occurred on a platform during a given window of time (e.g., ``weekly average number of instances of dehumanization, per user''). Correspondingly, a bridging metric could be defined to be the change in such a relation metric over time (e.g., ``change in the weekly average number of instances of dehumanization, per user''), or the overall prevalence of a motif thought to be bridging (``weekly average number of participations in a diverse approval motif, per user'').

\subsubsection{Formal Measures}

\begin{figure}[!ht]
        \centering                                                                               \includegraphics[width=0.9\textwidth]{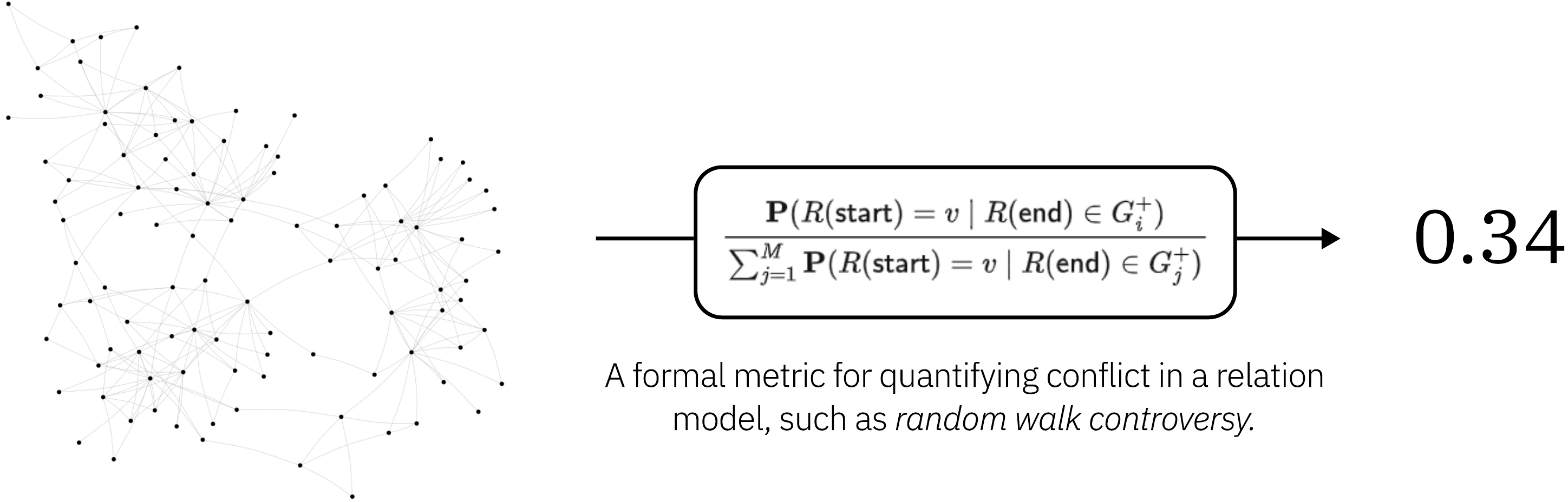}
        \vspace{5mm}
        \caption{\highlight{Example of a formal measure that could be used as a metric.} The formal measure used in this example is \textit{random walk controversy} \cite{interian2022}.}
        \label{fig:formal}
    \end{figure}

    Metrics can also be computed directly from a relation model, using summary statistics which are intended to capture normative judgments about whether the relations represented in the model are good, bad, improving or deteriorating. A visual intuition for this is given in Figure \ref{fig:formal}.

    Candidates for relation metrics include existing formal measures of polarization, which have been proposed for both space-based \cite{duclos2004,baumann2021,macy2021} and graph-based \cite{musco2021,matakos2017} models. However, there are many different kinds of polarization \cite{bramson2017}, and it is not clear whether existing polarization measures are appropriate targets for optimization. We believe the space of relation metrics is significantly under-explored, and have begun a survey of potential metrics at \href{https://bridging.systems/metrics/}{\url{https://bridging.systems/metrics/}} \cite{metrics2023}.
    
    \begin{exampleRecommender}[Relation Metrics]
        In the context of recommender systems and social media, relation metrics have mostly been proposed as summaries of graph-based models, such as follow networks on Twitter or the hyperlink networks of online news publications. These include metrics based on \newterm{homophily} (the degree to which nodes are connected to nodes that are similar to themselves), \newterm{modularity} (the number of intra-group connections relative to the number of inter-group connections), \newterm{random walk controversy} (a measure of how likely you are to cross between groups when randomly traversing the network), and \newterm{balance theory} (which measures how consistent the network is with properties such as ``my friend's friend is my friend'' and ``my friend's enemy is my enemy''). See  \textcite{interian2022} for a comprehensive review.
    \end{exampleRecommender} \vspace{-5mm}

    \begin{questions}[Metrics]

        \item How can we evaluate the quality of relation metrics and bridging metrics in a given context? What are their relative strengths and weaknesses?

        \item Given the challenges of measuring the effects of online platforms \cite{thorburn2022b}, how can we produce bridging metrics that can be interpreted as causal estimates?
        
        \item How might bridging metrics be misaligned, or be affected by the different versions of Goodhart's Law \cite{manheim2018}?
        
        \item What can be done to minimize negative side effects if relation metrics are used as targets for optimization?

        \item What is a reasonable single gold standard metric for bridging, that can provide useful intuition---a ``GDP of bridging''?

    \end{questions}

\subsection{Evaluation}
\label{sec:evaluation}

    The metrics described above (and by extension, their component signals) are intended to be optimized via changes in the design of an attention allocator. For this reason, it is important that these formalisms are closely connected to the underlying bridging goal (an increase in mutual understanding and trust across divides, creating space for productive conflict, deliberation, or cooperation). If the metrics do not capture the aspects of reality we care about, then efforts to optimize them will be at best ineffective, and at worst harmful. %

    Fortunately, there is a mature literature in the field of \textit{measurement theory} that provides a framework for thinking about the quality of relation metrics and bridging metrics. In this section we discuss two properties that it is good for metrics to have---validity and reliability---and consider what they mean in the context of bridging systems.

\subsubsection{Validity}
\label{sec:validity}

    Validity refers to whether a metric actually measures what it claims to measure. More precisely, validity is the ``degree to which evidence and theory support the interpretations of [metrics] for proposed uses of [those metrics]'' \cite{aera2014}.

    To see the importance of validity for relation and bridging metrics, consider the consequences of optimizing for a metric that evidence suggests has poor validity. For example, a platform with good intentions may decide to reward exposure diversity in their recommender system---that is, to specifically show their users content from political leanings they disagree with. They might be ``successful'' according to this measure, and interpret the increasing diversity of content viewed as indicative of increasing mutual understanding and trust across divides. However, it is possible that increased exposure to contrary viewpoints would actually cause people to double down on their existing beliefs and hence cause relations to deteriorate \cite{bail2018}. The reason for the discrepancy between the platform's good intentions and the regrettable outcome is that exposure diversity has poor validity as a measure of bridging.

    Relation and bridging metrics may lack validity for several reasons. They may be based on misconceptions (as in the case of exposure diversity). They may track the quality of relationships in a narrow scope observable by the platform, but not correlate with the quality of relationships in broader society. They may identify content that is bridging if shown to one person, but not bridging when shown to 10 million (due to second-order effects). They may be based on relation models that are not expressive enough to capture the underlying plurality of perspectives (this may be the case with space-based models \cite{bogomolnaia2007,thorburn2022d}). Or they may succumb to Goodhart's Law \cite{manheim2018}---that is, they may be inflated by actions that do not respect the original spirit of the metric. For example, promoting puppy videos may create more cases of diverse approval, but not have any effect on the quality of political discourse.

    Documenting design choices related to relation and bridging metrics, such as in reward reports \cite{gilbert2022}, might help monitor efforts to improve validity over time.

\subsubsection{Reliability}

    Reliability is the degree to which a metric takes on similar values when calculated in similar contexts. A reliable relation metric will give similar values if calculated multiple times in quick succession, if independently calculated by different teams within a company, or if calculated on multiple random samples of a population of platform users.
    
    Reliability is important because it is a prerequisite for validity. If a metric is too noisy or cannot be replicated, then it cannot be valid as an optimization target.

    Relation and bridging metrics may lack reliability (of various sorts), for several reasons. They may be easily influenced by ephemeral current events such as political scandals. They may be affected by ongoing changes in the design or affordances of a platform, or in changes to the composition of its user base, that undermine the ability to interpret changes in the value of the metric across time. They may depend on an automated classifier which is iteratively being updated or improved, effectively changing the metric that is calculated. They may depend on survey questions which, when translated into different languages, lead to response distributions that are no longer comparable.

    \begin{questions}[Evaluation]
        
        \item What are the limitations of different specific relation models (see e.g., \cite{thorburn2022d}), signals, or metrics intended for bridging? How significant are they?
        
        \item What gold standard measures or benchmarks should we use to evaluate relation models, signals, and metrics intended for bridging? 
        
        \item Which signals and metrics intended to promote bridging are more resistant to strategic manipulation?
        
    \end{questions}

\clearpage

\section{Discussion}
\label{sec:discussion}

    In this paper we have emphasized connections and open questions across three domains---recommender systems, collective response systems, and human facilitation---to help formalize an interdisciplinary research area focused on bridging. 

\subsection{Implementation}

    Table \ref{table:bridging} provides a summary of concrete steps that can be taken by practitioners to incorporate bridging into attention allocators.

    \begin{table}[H]
        \sffamily\centering
        \begingroup
        \renewcommand{\arraystretch}{1.5}
        \begin{tabular}{@{}lp{12.6cm}@{}}
        \toprule
        \textbf{Allocation Process}
                    & Include bridging impact(s) as a predicted impact \\
                    & Use bridging signals---give meaningful weight to bridging impact(s) in the value model \\
        [3pt]\midrule
        \textbf{Learning Process}
                    & Collect / retrieve / elicit data on bridging signals and relations \\
                    & Maintain and update a relation model \\
                    & Maintain and update model(s) that predict the impact of allocations on the relation model \\
        [3pt]\midrule
        \textbf{System Design}
                    & Monitor both relation and bridging metrics over time \\
                    & Consider bridging metrics when interpreting the outcomes of A/B tests \\ 
                    & Collect data that will help you do bridging better in the future \\
        [2pt]\bottomrule
        \end{tabular}
        \endgroup
        \caption{Examples of how to incorporate bridging into attention allocators.}
        \label{table:bridging}
    \end{table}

\subsection{Challenges, Limitations, and Risks}

    The idea of bridging systems and the interdisciplinary research we aim to articulate are not free from challenges or controversy, and should not be considered a panacea. Here we explore these potential challenges and related limitations and risks. After grounding the discussion in terminology and common misunderstandings we articulate challenges to the goal of bridging itself, to our formalization of bridging, and to the implementation of bridging. 

\paragraph{Intuition versus formalism}

    We defined the (intuitive) bridging goal as follows: \emph{an increase in mutual understanding and trust across divides, creating space for productive conflict, deliberation, or cooperation}. It is common when going from intuitions to formalisms that only a small part of the intuition is captured, sometimes with significant negative societal implications. We attempt to navigate that dilemma by making the intuition explicit, and emphasizing it as the ultimate goal to measure success (and formalisms) against. We relatedly don't attempt to make the formalism fully capture this intuition, as no single formalism understandable by a human is likely to do so. We expect that the best outcomes are likely to result from relying on a broad set of bridging metrics that address different aspects of the intuition.\footnote{We also recognize that this level of investment will not always be feasible. Even creating a set of modular bridging metrics and algorithms appropriate to a large variety of contexts is a significant public goods challenge.}

\paragraph{Common misunderstandings}

\begin{figure}[!ht]
        \centering                                                                      \includegraphics[width=.9\textwidth]{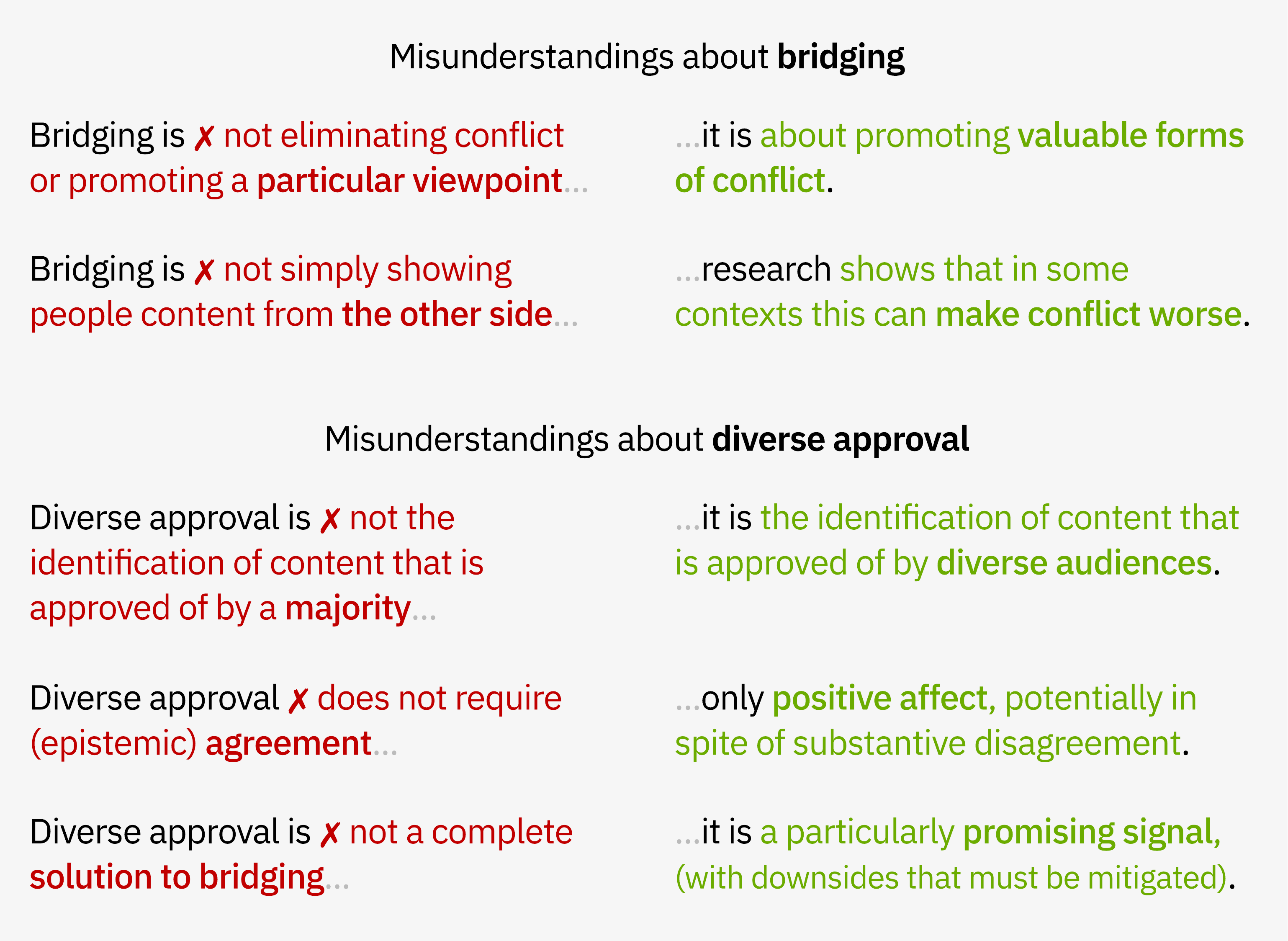}
        \caption{\highlight{Common misunderstandings.} Five common misunderstandings about bridging and diverse approval.}
        \label{fig:misconceptions}
    \end{figure}

    This notion of bridging is commonly misunderstood (and may be incorrectly formalized) in ways that would lead to outcomes that fall outside of the definition. Bridging does \textit{not} mean bringing everyone to a homogeneous center or eliminating conflict. Pluralism and productive conflict are valuable. Bridging also does \textit{not} mean just showing content from across divides; research suggests that this can sometimes \textit{increase} division \cite{bail2021}. Additional common misunderstandings are listed in Figure \ref{fig:misconceptions}.

\paragraph{Challenges to the goal of bridging}

    As the bridging intuition is imprecise, it can have different meanings at different levels of abstraction, formalism, and implementation. This makes the notion of bridging and the formalisms used to measure or apply it contestable. To give a powerful example, one study has found that there is an intervention which simultaneously \textit{reduces} partisan animosity and \textit{increases} support for partisan violence, as measured by different survey questions \cite{voelkel2022}. The decision of whether this constitutes a net improvement or deterioration---and more generally, the decision of how to trade off different interpretations or formalizations of bridging---may be both statistically technical and significantly normative.

    Many would argue that there should not even be an attempt to bridge some kinds of divides due to the abhorrence of particular groups, or their lack of respect for human rights \cite{margalit2009}. This implies a key question: when is bridging appropriate, and when is it not \cite{powell2022}? There are also arguments that destructive conflict may be an important driver of social change, so reducing destructive conflict could limit the potential for such change \cite{coser1998,mouffe1999,kreiss2023}. It has been argued that bridging could ``interfere with important processes that may be necessary for a democracy to determine the best path forward'' \cite{clyde2022}.

    It is also possible that bridging is too neutral a goal, in that bridges can form around ideologies or opinions that are considered harmful. For example, there are likely some people on the political left and political right who share a prejudice towards a particularly disadvantaged minority group. Such a prejudice bridges those on the left and right, but is not a desirable form of bridging. For this reason, one may need to specify additional constraints on the kinds of bridging that should be facilitated. Relatedly, in contexts where other factors, such as the quality of information, are far more important than mutual understanding (and where these are not as correlated as they appear to be with Twitter's Community Notes \cite{wojcik2022}), the bridging goal may also be too neutral (at least on its own).

    Finally, there may be differential benefits of bridging to particular groups. For example, some groups may be less willing or able to change than others, perhaps due to significant punishments for engaging outside of the group dogma or because some groups do much more ``internal bridging work'' than others.

\paragraph{Challenges to our formalization of bridging}

    Formalizing bridging around attention allocation does not account for material sources of division, such as physical violence, economic inequality, or historical injustices. Extensions of bridging to more general allocation problems (e.g., around physical resources) may perhaps be helpful in addressing some of these critiques, but bridging would still be far from a panacea. In a world with many divides, attention allocations that bridge some divides may exacerbate others, and formalisms may not be able to meaningfully balance those contradictory impacts.
    
    Some may argue that any formalism or optimization of concepts as complex and interpersonal as bridging is inherently inhumane or invalid. Regardless of whether humane formalization of bridging is possible, it is true that any practical formalism will be a considerable simplification of the complex psychological and social changes that correspond to the bridging goal. Such simplifications may not reflect important aspects of bridging and would benefit from ongoing critique and improvement.\footnote{If AI advances enable simulation and automation of high-quality human-like facilitation, that may allow us to move beyond formalisms, though at the risk of reduced understanding of what such systems are doing.}
    
    Specific choices of relation models, relation metrics, and bridging metrics may lack validity (see Section \ref{sec:validity}) or may not be aligned with other societal values and goals. Specific approaches might cause unintended side effects or consequences, some of which may be significant. Some argue that bridging constitutes an attempt to influence a population, and that it is an open question whether this can be done in a manner that respects human agency and avoids unwanted manipulation \cite{bakcoleman2021,carroll2022,carroll2023} (though a similar argument might hold for any system that allocates attention). At the very least, bridging can be viewed as a form of ``intrinsic governance'' \cite{lazar2023a} or ``social mediation'' \cite{reisman2023}: it is an attempt to make some types of communicative acts easier and others more difficult. There are significant, unresolved questions regarding the contexts in which such governance decisions are ethically justified \cite{lazar2023a,lazar2023b}, and how they can be made in a way that appropriately reflects ideals of democracy and subsidiarity---especially in light of a status quo where such decisions are already frequently made with far less benevolent aims. %

\paragraph{Challenges to a technological implementation of bridging}

    Adversaries are likely to try to use fake or manipulated accounts in order to influence bridging-based ranking algorithms. This is already true of existing algorithms and similar safeguards may apply, but it is possible that there are either more or less algorithmic vulnerabilities as a result of a bridging-based ranking approach. We may also not have the technological capacity to predict societal outcomes with sufficient accuracy to implement important aspects of the intuition for bridging \cite{narayanan2019}. Instead, we may have to rely on heuristic signals like diverse approval (Section \ref{sec:signal-examples}), which then need to be either validated or justified as having impacts ``close enough'' to the bridging goal (Section \ref{sec:evaluation}). %

    By including bridging in a value model, there is a trade-off between bridging and other goals, such as relevance or engagement. Including bridging may, in some cases, reduce the extent to which these other goals can be achieved. Bridging is less likely to be implemented by large social media and search engine companies unless it is either economically beneficial, neutral, or there is significant and sustained external pressure (or regulation). While it is possible that bridging-based ranking may reduce moderation costs and increase long-term user retention, it may have short-term negative impacts on engagement and growth. Bridging may be particularly challenging to maintain in an environment with sustained competition between platforms for attention, and easiest to maintain in a monopoly. However, in spite of these challenges, there have now been publicly disclosed implementations of bridging across multiple major platforms, albeit in limited areas thus far, and we have been informed of others that have not yet been made public \cite{twitter2022a,facebook2022,facebook2022a,facebook2022b}.
    
    There is considerable uncertainty over the impacts of a bridging implementation. It is possible that the creation of bridging content cannot be incentivized or that a particular implementation of bridging practically results in predominantly benign, widely acceptable content such as cat videos being promoted. Some may argue that algorithmic attention allocators are delegated too small a portion of human attention (relative to, say, interpersonal relationships or mainstream media organizations) for efforts incorporating bridging to have any meaningful impact. Bridging-based ranking may also be less impactful or less tractable to implement within distributed or decentralized social media environments, or in models of social media and mass communication that are yet to emerge (e.g., those involving federation or middleware) and in models that may not be based on ordinal ranking (e.g., those built around AI-generated summaries).

    Finally, technologies designed to influence the types of conflict in society are inevitably dual use. If we can use technology to make conflict better, we can also make it worse. It is possible that the recommender systems currently used by large social platforms already cause conflict to deteriorate as a side effect, but none are explicitly designed to ``increase polarization''. Developing technology for bridging may only be beneficial if the risks and impacts of such perverse misuse are sufficiently low.

\paragraph{Overcoming these challenges}

    A large part of the research required in this area is to better understand the extent to which these challenges can be avoided, mitigated, or managed. They must also be considered in the context of the status quo, key elements of which (e.g., optimizing for engagement in social media, political messaging and advertising) are rarely subject to the same level of caution or scrutiny before implementation that we are applying here. In other words, the risks of intervention must be balanced with the risks of inaction; and as we argue in more detail in \cite{ovadya2022}, the status quo for recommender systems leaves much to be desired.

    \begin{questions}[Challenges, Limitations, and Risks]
        \item How can we best overcome the most critical challenges, limitations, and risks?
        \item For what contexts and under what conditions is bridging appropriate, and where is incorporating bridging worse than the status quo? 
        \item Are there specific additional objectives that need to be incorporated in order for bridging to be helpful in some contexts, and if so, what are they and how can they be operationalized?
        \item To what extent is the bridging goal in tension with other goals such as engagement or relevance?
        \item Are there relatively under-explored approaches to bridging that might address many of these challenges? For example, a system that identifies signals of perception gaps \cite{yudkin2019}, and dynamically elicits (or generates) content or activities that can help reduce those gaps; potentially all within an existing system such as a newsfeed recommender or a language model chatbot.

    \end{questions}

\clearpage

\section{Conclusion}

    Our social spaces should not \emph{default to divisive}. Bridging is a core part of healthy social fabrics and the systems that allocate attention within them, whether human or machine, implicit or explicit. Without sufficient bridging, destructive conflict may undermine our relationships and prevent us from cooperating effectively to respond to societal challenges.

    While bridging may help counteract the ``bias toward division'', it is unlikely to address the myriad of other biases that social spaces may entrench. We see this work as complementary to the lines of research around other biases, and it is important that we also keep in mind (and measure where appropriate) these other forms of bias when designing for bridging. We should also caveat that designing systems for meaningful positive interactions is part art, and in this paper we have framed it as a science. We encourage work that can go beyond ``shallow bridging heuristics'' to ``deep bridging''---from bonds around cat videos to bonds about our very different struggles in living complex lives. In this paper we have also focused primarily on bridging as it relates to the selection of what to attend to among a set of options, but the intuitions and formalisms we have explored can also be adapted to the \emph{synthesis} of such options---i.e., to generative systems and foundation models such as ChatGPT and GPT-4 \cite{bakker2022FinetuningLanguageModelsa, brown2020LanguageModelsArea}. 

    New kinds of social spaces, both online or offline, do not always incorporate hard-won lessons from the old. Modern computational and communication technologies have changed the structure of civilization, bringing billions of people onto the internet \cite{bakcoleman2021}. But as we moved online, bridging is also one of the elements of offline spaces that we have most under-resourced.

    Belatedly, some online spaces are catching up and setting an example. The deployment of the Community Notes feature on Twitter \cite{twitter2022,wojcik2022} accelerated the adoption of bridging-based ranking and demonstrated that signals like diverse approval hold significant promise even at significant scale. Since the publication of the original bridging-based ranking paper \cite{ovadya2022}, we have heard from people interested in bridging across organizations of all sizes and kinds, from the largest tech companies, to recently incorporated startups, to the technology teams at traditional media organizations. Through this paper, we have attempted to provide a framework for understanding how such bridging systems can be understood and developed, and laid out a set of research questions to support and accelerate their safe deployment.

    There remains the normative question of to what extent bridging is appropriate for a given system, which is outside of the scope of this paper. Ideally, such decisions would be a collective choice of those impacted by the system. While such collective decision-making might seem impractical for, say, platforms with billions of users, representative deliberations such as citizens' assemblies, in combination with collective response systems, provide a potentially viable mechanism for such democratic decision-making at global scale \cite{ovadya2023, ovadyaMetaRanGiant, ovadya2021}.
    
    Our omnipresent social media, search, and synthetic media systems were not designed to satisfy the bridging goal: to increase mutual understanding and trust across divides, creating space for productive conflict, deliberation, or cooperation. But these systems that help allocate our attention are not irredeemable---they have the potential to support a more deliberative, peaceful, and pluralistic future.

\clearpage

\section*{Acknowledgements}

    We thank Tim van Gelder and Colin Megill for sharing information and insights from their work on YourView and Polis, respectively, and members of the GETTING-Plurality Research Network at Harvard University (\href{https://gettingplurality.org}{\url{gettingplurality.org}}) for their ongoing feedback and support. We would additionally like to thank Danielle Allen, Natania Antler, Katy Glenn Bass, Jay Baxter, Priyanjana Bengani, Leisel Bogan, Paul Bouchaud, Jason Burton, Micah Carroll, Alan Chan, Austin Clyde, Madeleine Daepp, Fernando Diaz, Joe Edelman, Davis Foote, Thomas Gilbert, Sam Hinds, Ravi Iyer, Amritha Jayanti, Julia Kamin, Emillie van de Keulenaar, Andrew Konya, David Krueger, Stephen Larrick, Seth Lazar, Jesse McCrosky, James Mickens, Smitha Milli, Arvind Narayanan, Max Nickel, Kyle van Oosterum, Kathy Pham, Maria Polukarov, Helena Puig Larrauri, Joaquin Quiñonero Candela, Richard Reisman, Afsaneh Rigot, Bruce Schneier, Christopher Small, Jonathan Stray, Ted Suzman, Carmine Ventre, Jessica Yu, Glen Weyl and Cathy Wu, among others, for helpful discussions and feedback. Any errors or limitations of this work remain those of the authors.

    Aviv Ovadya was supported in part by a Technology and Public Purpose Fellowship at the Belfer Center for Science and International Affairs, Harvard Kennedy School. Luke Thorburn was supported by UK Research and Innovation [grant number EP/S023356/1], in the UKRI Centre for Doctoral Training in Safe and Trusted Artificial Intelligence (\href{https://safeandtrustedai.org}{\url{safeandtrustedai.org}}), King's College London. Graphics were developed in part with support from Tereza Flídrová through a grant from the Plurality Institute (\href{https://plurality.institute}{\url{plurality.institute}}).

\printbibliography
\clearpage

\end{document}